\documentstyle[12pt,epsfig]{article}
\input{epsf}
\newcommand{\be}{\begin{equation}}
\newcommand{\ee}{\end{equation}}
\newcommand{\bea}{\begin{eqnarray}}
\newcommand{\eea}{\end{eqnarray}}

\newcommand{\bk}{{\bf k}}

\newcommand{\vk}{{\rm {\bf k}}}
\newcommand{\vq}{{\rm {\bf q}}}

\def\la{\mathrel{\mathpalette\fun <}}

\def\fun#1#2{\lower3.6pt\vbox{\baselineskip0pt\lineskip.9pt
\ialign{$\mathsurround=0pt#1\hfil##\hfil$\crcr#2\crcr\sim\crcr}}}

\begin{document}

\title{Radiative decays of  basic scalar, vector and tensor
mesons and the determination of the
$P$-wave $q \bar q $ multiplet}
\author{A.V. Anisovich, V.V. Anisovich and V.A. Nikonov}
\date{\today}
\maketitle

\begin{abstract}

We perform simultaneous calculations of
the radiative decays of scalar mesons $f_0(980)\to \gamma\gamma$,
$a_0(980)\to \gamma\gamma$,
vector meson
$\phi(1020)\to \gamma f_0(980)$, $\gamma a_0(980)$,
$\gamma \pi^0$, $\gamma \eta$, $\gamma \eta'$
and tensor mesons
$a_2(1320)\to \gamma\gamma$, $f_2(1270)\to \gamma\gamma$,
$f_2(1525)\to \gamma\gamma$  assuming  all these states
to be dominantly the $q\bar q$ ones. A good description of the
considered radiative decays is reached by using
almost the same radial wave
functions for  scalar and tensor mesons that supports the idea
for the $f_0(980)$, $a_0(980)$ and
$a_2(1320)$, $f_2(1270)$, $f_2(1525)$ to belong
to the same $P$-wave $q\bar q$ multiplet.

\end{abstract}

\section{Introduction}

Despite a long history of the $P$-wave $q\bar q$
multiplet \cite{Gatto} the problem of definition of $q\bar q $
scalars is still a subject of lively discussion, see e.g.
\cite{klempt,montanet,petry} and references therein. Radiative decays
of mesons may serve as a useful tool for the study of $q\bar q$
structure of mesons, in particular, $P$-wave
$q\bar q$ component in
$f_0(980)$ and $a_0(980)$. In this way, it is rather important to
investigate simultaneously the other mesons which belong to the $P$-wave
$q\bar q$ multiplet, namely,  tensor mesons
$a_2(1320)$, $f_2(1270)$ and $f_2(1525)$. In the present paper,
 combined calculations of the decays
$a_0(980)\to \gamma\gamma$,
$f_0(980)\to \gamma\gamma$,  $a_2(1320)\to \gamma\gamma$,
$f_2(1270)\to \gamma\gamma$, and
$f_2(1525)\to \gamma\gamma$ are carried out assuming the
$q\bar q$ radial wave functions in these mesons to be nearly the same.

Radiative decays of the $\phi$-meson are another source of important
information on scalar mesons. We have calculated the decay processes
with the production of mesons belonging to scalar and
pseudoscalar sectors:
$\phi(1020)\to \gamma f_0(980), \gamma a_0(980)$ and
$\phi(1020)\to \gamma \pi_0, \gamma \eta, \gamma \eta'$. These
latter, of the  type of $V\to \gamma P$, are the classical
reactions, which had been used rather long ago for the determination of
 $q\bar q$ structure of vector and pseudoscalar mesons \cite{pnpi}.

We believe that
simultaneous description of the processes $S\to \gamma\gamma $,
$T\to \gamma\gamma $,
$V\to \gamma S$ and $V\to \gamma P$
is a necessary test for the whole calculation procedure
and determination of the $P$-wave $q\bar q$ multiplet.

In calculations of the decay form factors we
use spectral integration over $q\bar q$ states together with
the light-cone wave functions for the $q\bar q$ mesons;
the method of the spectral integration for the form
factor amplitudes
has been developed in a set of papers \cite{AKMS,pi,eta,f0gg}.

In Section 2 we present necessary elements of the
technique for the calculation of radiative decay amplitudes.
The detailed presentation of the
technique for the description of composite $q\bar q$ systems
can be found in
Refs. \cite{pi,eta,f0gg}, where the pion form factor was studied
together with transition form factors $\pi^0 \to \gamma (Q^2)\gamma $,
$\eta \to  \gamma (Q^2)\gamma$ and $\eta' \to  \gamma
(Q^2)\gamma$.  The method of spectral integration works for the form
factor amplitudes which obey the requirement of analyticity, causality
and gauge invariance.  The used technique allows one to  introduce the
composite-particle wave functions and perform calculations in terms of
the light-cone variables.

It is worth noting that this calculation technique for the processes
involving bound states has a broader applicability than for the $q\bar
q$ systems only: in \cite{AKMS} this method got its approbation by
describing the deuteron as a composite $np$ system, then this very
technique was applied to heavy mesons \cite{melikhov}.
For the convenience of a reader, in Section 2 we recall briefly the
basic points of this approach. Then we give necessary formulae for
the calculation of  partial widths for the decays
$V\to \gamma S$, $V\to \gamma P$, $S\to \gamma\gamma$
and $T\to \gamma\gamma$.

Results of the calculation are presented
in Section 3. First, we discuss the decay
$\phi(1020) \to \gamma f_0(980)$.
Our calculations show that the data on branching ratio
$BR(\phi (980) \to \gamma f_0(980))
= (3.4 \pm 0.4 ^{+1.5}_{- 0.5})\cdot 10^{-4}$ \cite{novosib,PDG-00}
may be described  assuming
 the $q\bar q$ structure of $f_0(980)$
and varying  $s\bar s $ and $n\bar n=(u\bar u+d\bar d)/\sqrt 2 $
components in a broad interval. For the flavour wave function written
as
$\psi_{flavour}[f_0(980)]= n\bar n \cos\varphi+s\bar s \sin\varphi $
the agreement with data
is reached with $ 25^\circ \le |\varphi|\le 90^\circ $.

In Section 3, we calculate also  partial widths for the
decays $\phi(1020) \to \gamma \eta,\gamma \eta',\gamma \pi^0$,
with the same technique as has been used for the
reaction $\phi(1020) \to \gamma f_0(980)$ and with the same
parametrization of the $\phi$-meson wave function. The calculations
demonstrate a good agreement with  data
as well. It should be stressed that in fact the decays $\phi(1020) \to
\gamma \eta,\gamma \eta'$ are calculated without any free parameter:
  these decays are governed by $s\bar s$ components in
$\eta$ and $\eta'$ which are well known; the wave functions of the
basic pseudoscalar and vector mesons are also known, see e.g.
\cite{pi,eta,ABMN}. So, the calculations of the
decays $\phi(1020) \to \gamma \eta,\gamma \eta'$ are needed for
the verification of the method only, and the results
provide us with a strong argument that the applied method
for calculation of radiative decays of $q\bar q$ mesons is wholly
reliable. The decay $\phi(1020) \to \gamma \pi^0$ allows us to
estimate the admixture of the $n\bar n$ component in $\phi (1020)$.
With the flavour wave function of the $\phi (1020)$ written as
$\psi_{flavour}(\phi (1020))= s\bar s \cos\varphi_V+n\bar n
\sin\varphi_V $, we have $ |\varphi_V|\le 4^\circ $.
Partial width of $\phi(1020) \to \gamma a_0(980)$ is also
proportional to the probability of  $n\bar n$ component in the $\phi
(1020)$; we discuss this decay in Section 3 as well.

Two-photon radiative decays provide  important
information about $P$-wave $q\bar q$ mesons; the technique for the
calculation of  scalar- and tensor-meson decays, $S\to \gamma\gamma$
and $T\to \gamma\gamma$, is presented in Section 2.

Under the assumption of
the $q\bar q$ structure of $a_0(980)$ and $f_0(980)$,
analysis of  partial widths $a_0(980)\to \gamma\gamma$ and
$f_0(980)\to \gamma\gamma$ has been performed in \cite{f0gg}. The data
for $a_0(980)\to \gamma\gamma$ are in a reasonable agreement with
calculation. Concerning the $f_0(980)$, the extraction of the signal
$f_0(980)\to\gamma\gamma$ from the measured spectra
$\gamma\gamma\to\pi\pi$ faces  strong interference  "resonance +
background", thus
resulting in uncontrollable errors (see, for example,
the K-matrix calculation of the $S$-wave spectra
$\gamma\gamma\to\pi\pi$ \cite{AA}).  The recently
obtained partial width $\Gamma(f_0(980)\to \gamma\gamma) = 0.28 ^{+
0.09}_{-0.13}$ keV \cite{pennington} is by a factor 2 smaller than the
averaged value reported previously ($0.56 \pm 0.11$ keV \cite{PDG-98}).
In Section 3 we re-analyse the decay $f_0(980)\to \gamma\gamma$ using
new data for the width. Thus we get two allowed intervals for
the $n\bar n/s\bar s$ mixing angle:  $ 80^\circ \le \varphi \le
93^\circ $ and $ (-54^\circ)\le \varphi \le (-42^\circ) $.  The
restrictions for  mixing angle $\varphi$ which come from the
combined analysis of radiative decays $\phi(1020)\to\gamma f_0(980)$
and $f_0(980)\to\gamma\gamma$ are discussed in Section 4, we have two
solutions for $ \varphi$:
\begin{equation}
\varphi=-48^\circ \pm
6^\circ , \qquad \varphi=86^\circ \pm 3^\circ \; .
\end{equation}
For positive mixing angle, the analysis gives us strong restriction
for the value of radius of the $f_0(980)$: $ R^2_{f_0(980)} \le 7 \;
{\rm GeV}^{-2} $ (remind,  the pion radius squared is $ R^2_\pi \simeq
10 \; {\rm GeV}^{-2} $).

In Section 3 we discuss also the results of the  calculation of
 tensor-meson two-photon decays:  $a_2(1320)\to \gamma\gamma$,
$f_2(1270)\to \gamma\gamma$ and $f_2(1525)\to \gamma\gamma$. The form
factors of corresponding transitions depend strongly on the choice
of the vertex $T\to q\bar q$. In  line with the $q\bar q$
classification, we perform comparison with data for the vertex which
is related to  dominant $q\bar q$ $P$-wave. The results are in
reasonable agreement with the measured partial width $\Gamma
(a_2(1320)\to \gamma\gamma)$,
calculations being carried out with the wave function of $a_2(1320)$
whose characteristics are close to  those of $a_0(980)$, namely,
 $ R^2_{a_2(1320)} \simeq  R^2_{a_0(980)}
\simeq  7-12\; {\rm GeV}^{-2} $. Description of the two-photon decays
of $f_2(1270)$ and $f_2(1525)$ fixes $n\bar n /s\bar s$ ratio for these
states.  With  flavour wave functions written as
$\psi_{flavour}[f_2(1270)] = n\bar n \cos\varphi_T +s\bar s
\sin\varphi_T $
and
$\psi_{flavour}[f_2(1525)] = - n\bar n \sin\varphi_T   +s\bar s
\cos\varphi_T $,
simultaneous description of the data can be reached at
$ R^2_{f_2(1270)} \simeq  R^2_{f_2(1525)}
\simeq  7-10\; {\rm GeV}^{-2} $  requiring either $\varphi_T \simeq
0^\circ $ or $\varphi_T\simeq 25^\circ $.

Simultaneous description of  radiative decays of
scalar and tensor mesons,
$f_0(980)$, $a_0(980)$ and
$a_2(1320)$, $f_2(1270)$, $f_2(1525)$,
with the use of  similar radial wave functions, argues in favour of
their belonging to the same $P$-wave $q\bar q$ multiplet.

\section{Radiative decays in the framework of spectral integration
technique}

Given here are the formulae for partial widths of the
radiative decays:
$V\to \gamma S$, $V\to \gamma P$, $S\to \gamma\gamma$
and $T\to \gamma\gamma$. Using as an example  the reaction
$V\to \gamma S$,
we present  necessary elements of the spectral integration technique
applied for the calculation of  transition form factors.

\subsection{Moment-operators for the transition amplitudes $S\to
\gamma\gamma$, $T\to \gamma\gamma$, $V\to \gamma S$ and $V\to \gamma
P$ }

For amplitudes under consideration, we present the moment operators
 for two-photon decays of the scalar
and tensor mesons: $f_0(980)$, $a_0(980)\to\gamma\gamma$ and
$f_2(1270)$, $a_2(1320)$, $f_2(1525)\to\gamma\gamma$, and
for radiative decays of the $\phi$-meson:
$\phi(1020)\to\gamma\pi^0$, $\gamma\eta$, $\gamma\eta'$,
$\gamma a_0(980)$, $\gamma f_0(980)$.
Systematic presentation of the moment operators is given in
\cite{moment}.

\subsubsection{Transition amplitude $S\to \gamma \gamma $}
The transition amplitude
$S\to \gamma_\perp (q^2)\gamma_\perp  (q'^2)$
for the transversely polarized photons
reads:
\be
 A_{\alpha\beta}= e^2 F_{S\to \gamma \gamma}(q^2,q'^2)
g^{\perp\perp}_{\alpha\beta}\ .
\label{1.6}
\ee
Here $e$ is the electron charge ($e^2 /4\pi =\alpha = 1/137$);
the indices
$\alpha$,$\beta$  refer to the photons;
$q$ and $q'$ are the photon momenta.
The metric tensor $g^{\perp\perp}_{\alpha\beta}$
works in the
space orthogonal to $p=q+q'$ and $q$:
\be
g^{\perp\perp}_{\alpha\beta}=g_{\alpha\beta}-
\frac{q_{\perp\alpha} q_{\perp\beta}}{q_\perp^2} -
\frac{p_\alpha p_\beta}{p^2}\; ,\qquad
q_{\perp \alpha}= g^{\perp}_{\alpha\alpha'} q_{\alpha'}\; ,\qquad
g^{\perp}_{\alpha\alpha'}=g_{\alpha\alpha'}-
\frac{p_\alpha p_{\alpha'}}{p^2}\; .
\label{2.4}
\ee

\subsubsection{Tensor meson decay amplitude $T\to \gamma\gamma$ }

The $T\to \gamma\gamma$ decay amplitude has
the following structure:
\be
A_{\mu\nu,\alpha\beta}= e^2 \left [
S^{(0)}_{\mu\nu,\alpha\beta} (p,q)
 \; F^{(0)}_{T\to \gamma\gamma}(0,0) +
S^{(2)}_{\mu\nu,\alpha\beta} (p,q)
 \; F^{(2)}_{T\to \gamma\gamma}(0,0)
\right ]\, .
\label{1.7}
\ee
where $S^{(0)}_{\mu\nu,\alpha\beta} (p,q)$ and
$S^{(2)}_{\mu\nu,\alpha\beta} (p,q)$ are the moment operators, indices
$\alpha,\beta$ refer to photons and $\mu,\nu$ to the tensor meson.
Two transition form factors for the transversely polarized photons
$T \to \gamma_\perp (q^2)\gamma_\perp (q'^2)$,
namely, $ F^{(0)}_{T\to \gamma\gamma}(q^2 ,q'^2)$ and
$ F^{(2)}_{T\to \gamma\gamma}(q^2 ,q'^2)$,
depend on the photon momenta squared $q^2 $ and $q'^2 $;
the limit  $q^2 =q'^2 =0$  corresponds to the two-photon
decay.

The moment-operators read:
\be
S^{(0)}_{\mu\nu,\alpha\beta}(p,q)=
\left ( \frac{q_{\perp\mu} q_{\perp\nu}}{q_\perp^2}-\frac13
g^\perp_{\mu\nu} \right )
g^{\perp\perp}_{\alpha\beta}
\label{1.8}
\ee
and
\be
 S^{(2)}_{ \mu\nu\; ,\;\alpha \beta }(p,q)
=
g^{\perp\perp}_{\mu\alpha} g^{\perp\perp}_{\nu\beta} +
g^{\perp\perp}_{\mu\beta} g^{\perp\perp}_{\nu\alpha}
-g^{\perp\perp}_{\mu\nu}
g^{\perp\perp}_{\alpha\beta} \ .
\label{1.9}
\ee
The moment-operators are orthogonal in the space of photon
polarizations: \\ $S^{(0)}_{\mu\nu,\alpha\beta}
S^{(2)}_{\mu'\nu',\alpha\beta}=0$.

\subsubsection{Transition amplitude $V\to \gamma S$}

Transition amplitude $V\to \gamma_\perp (q^2) S$
for the transversely polarized photon
takes the form:
\be
 A_{\mu\alpha}=
 e F_{V\to \gamma S}(q^2) g^{\perp\perp}_{\mu\alpha}\; .
\label{1.4}
\ee
The index
$\alpha$ refers to the photon and $\mu$ to the vector meson;
$p$ and $q$ are the momenta of the initial vector meson and
photon. The metric tensor $g^{\perp\perp}_{\mu\alpha}$
works in the
space orthogonal to $p$ and $q$.
The limit $q^2 =0$ corresponds to the radiative decay of  vector
meson.

\subsubsection{ Transition amplitude $V\to \gamma P$}

The spin operator for
the amplitude $V\to \gamma P$ contains antisymmetric
tensor $\epsilon_{\mu\nu\alpha\beta}$, and
the amplitude has the following structure:
\be
A_{\mu\alpha}=e\; \epsilon_{\mu\alpha\nu_1\nu_2}p_{\nu_1} q_{\nu_2}
F_{V\to \gamma P}(q^2)\ .
\label{1.5}
\ee
The notations are the same as  for $V\to \gamma S$.

\subsection{Form factor for the radiative decay
$V\to\gamma_\perp (q^2)S$.}

The method of calculation of the three-point
form factor amplitudes in terms of the spectral representations
over the $q\bar q$ intermediate state masses was developed
in \cite{pi}.
Here we give schematic presentation of the method
using as an example the reaction
$V\to\gamma_\perp (q^2)S$.

\subsubsection{ Double spectral representation of the form
factor.}
Assuming the $q\bar q$ structure for the initial ($V$) and
final ($S$) mesons,
the amplitude of the decay $V\to \gamma S$ is determined
by the subprocesses
$V\to q\bar q$ and $ q\bar q\to S$, with the emission
of $\gamma(q^2)$, see Fig. 1$a$.
Corresponding three-point loop diagram
is calculated using double spectral
representation over intermediate
$q\bar q$ states: they are marked by dashed lines in Fig. 1$b$.

To be illustrative, let us start with the three-point
Feynman diagram. For the process of Fig. 1$a$ one has:
\be
A_{\mu\nu}^{({\rm Feynman})}
=\int \frac{d^4k}{i(2\pi)^4}\; G_V \;
\frac{Z^{(q\bar q)}_{V\to \gamma S} \; S^{(V\to\gamma  S)}_{\mu\nu}}
{(m^2-k_1^2)(m^2-k'^2_1) (m^2-k^2_2)}\; G_S\ .
\label{2.1}
\ee
Here
$k_1$, $k'_1$, $k_2$ are quark momenta,
$m$ is the quark mass, and $G_V$, $G_S$ are
quark-meson vertices; the quark charges are included into
 $Z^{(q\bar q)}_{V\to \gamma S}$.
The spin-dependent block reads:
\be
S^{(V \to\gamma  S)}_{\mu\nu}=-{\rm Sp} \left [
(\hat k'_1+m)\gamma^\perp_\mu (\hat k_1+m)\gamma^\perp_\nu
(-\hat k_2+m) \right ] \ ,
\label{2.2}
\ee
where the Dirac matrices
$\gamma^\perp_\mu $ and  $\gamma^\perp_\nu $ are orthogonal to the
emitted momenta:
$\gamma^\perp_\mu q_\mu =0 $ and $\gamma^\perp_\nu p_\nu =0$.

To transform the Feynman integral
(\ref{2.1}) into double spectral integral over invariant
$q\bar q$ masses squared,
one should make the following steps:\\
(i)  consider the corresponding energy-off-shell diagram, Fig. 1$b$,
with $P^2=(k_1+k_2)^2 \ge 4m^2$, $P'^2=(k'_1+k_2)^2 \ge 4m^2$ and
fixed momentum transfer squared $q^2 =(P-P')^2$,\\
(ii) extract the invariant amplitude by separating spin operators,\\
(iii) calculate the discontinuities of the invariant amplitude
over intermediate $q\bar q$ states marked in Fig. 1$b$ by dashed lines.

The double discontinuity is the integrand of the spectral
integral over $P^2 $ and $P'^2 $.
Furthermore, we put the following notations:
\be
P^2=s, \qquad P'^2=s' \ .
\label{2.3}
\ee
For the calculation of discontinuity, by cutting the Feynman diagram,
the pole terms of the propagators are replaced with
their residues:  $(m^2-k^2)^{-1} \to \delta (m^2-k^2)$. So, the
particles in the intermediate states
marked by dashed lines I and II in Fig. 1$b$
are mass-on-shell,
$k^2_1=k^2_2=k'^2_1=m^2$. As a result, the Feynman diagram
integration turns into the integration over phase spaces of
the cut states.
Corresponding phase space for the three-point diagram reads:
\be
d\Phi(P,P';k_1,k_2,k'_1) = d\Phi(P;k_1,k_2)d\Phi(P';k'_1,k'_2)
(2\pi)^3 2k_{20}\delta^{(3)}(\bk'_2-\bk_2) \ ,
\label{2.40}
\ee
where the invariant two-particle phase space
$d\Phi(P;k_1,k_2)$ is determined as follows:
\be
d\Phi(P;k_1,k_2) = \frac12 \frac{d^3k_1}{(2\pi)^3 2k_{10}}
\frac{d^3k_2}{(2\pi)^3 2k_{20}} (2\pi)^4 \delta^{(4)}(P-k_1-k_2) \ .
\label{2.5}
\ee
The last step is to single out the invariant component
from the spin factor (\ref{2.2}).
According to (\ref{1.4}), the spin factor (\ref{2.2}) is proportional
to the metric tensor,
$S^{(V \to\gamma  S)}_{\mu\nu} \sim g^{\perp\perp}_{\mu\nu} $,
which works in the space of the intermediate state momenta.
Then the spin factor $S^{{\rm tr}}_{V\to\gamma S}$, determined as
\be
S^{(V \to S)}_{\mu\nu} =
g^{\perp\perp}_{\mu\nu}S_{V\to\gamma S}(s,s', q^2),
\label{2.6}
\ee
is equal to:
\be
S_{V\to\gamma S}(s,s', q^2)=
-2m \left (4m^2 +s-s'+q^2
-\frac{4ss'\alpha(s,s',q^2) }{s+s'-q^2}\right ),
\label{2.8}
\ee
$$
\alpha(s,s',q^2)= \frac{q^2(s+s'-q^2)}{2q^2(s+s')-(s-s')^2-q^4}.
$$
Recall that, when going from (\ref{2.2}) to (\ref{2.8}), we use
the mass-on-shell relations
$(k_1k_2)=s/2-m^2$, $(k'_1k_2)=s'/2-m^2$, and $(k'_1k_1)=m^2-q^2/2$.

The spectral integration is carried out over the energy squared of
quarks in the intermediate states,
$s=P^2=(k_1+k_2)^2$ and $s'=P'^2=(k'_1+k_2)^2$, at fixed $q^2=(P'-P)^2$.
The spectral
representation for the amplitude $A_{V\to \gamma S}(q^2)$ reads:
\bea
\label{2.9}
A_{V\to \gamma S}(q^2)
&=&\int \limits_{4m^2}^\infty \frac{ds}{\pi} \int
\limits_{4m^2}^\infty \frac{ds'}{\pi}
\frac{G_V(s)}{s-M_V^2}\frac{G_S(s')}{s'-M_S^2}\times
 \\
&\times& \int d\Phi(P,P';k_1,k_2),k_1'\,
S^{{\rm (tr)}}_{V\to \gamma S}(s,s', q^2)
Z^{(q\bar q)}_{V\to\gamma S}\ . \nonumber
\eea
The spectral representation of the amplitude $A_{V\to \gamma S}(q^2)$
gives us the invariant part of $A_{\mu\nu}^{({\rm Feynman})}$,
Eq. (\ref{2.1}), when the vertices, $G_V(s)$ and $G_S(s')$,
 are constant.  Generally, the energy-dependent vertices can be
incorporated into spectral integrals. According to \cite{AKMS}, the
form factor of a composite system can be obtained by considering the
two-particle partial-wave  scattering amplitude $1+2\to 1+2$:
the pole sigularity of this amplitude corresponds to
the composite system.  The amplitude for the
emission of a photon by the two-particle-interaction system has two
poles related to the states "before" and "after"
electromagnetic interaction, and the two-pole residue of this amplitude
provides us the form factor of  composite system. When a
partial-wave scattering amplitude is treated using the
dispersion relation $N/D$-method, the vertex $G(s)$ is determined by
the $N$-function:  the vertex as well as $N$-function have
left-hand-side singularities which are determined
 by forces between the particles
1 and 2.

It is reasonable to name the ratios $G_V(s)/(s-m^2)$ and
$G_S(s')/(s'-m^2)$ the wave functions of vector and scalar particle,
respectively:
\be
\frac{G_V(s)}{s-m^2}=\psi_V(s), \qquad
\frac{G_S(s')}{s'-m^2}=\psi_S(s')\ .
\label{2.10}
\ee
Working with Eq. (\ref{2.9}) one can
express it in terms of the light-cone variables.

\subsubsection{ Light-cone variables}
One can transform Eq. (\ref{2.9}) to
the light-cone variables, using the boost along the $z$-axis.
Let us use the frame
where initial vector meson  moves along the
$z$-axis with the momentum
$p\to \infty$:
\be
P=(p+\frac {s}{2p},{\bf 0}, p), \qquad
P\; '=(p+\frac {s'+q^2_{\perp}}{2p}, -\vq_{\perp}, p).
\label{2.11}
\ee
In this frame the two-particle phase space is equal to
\bea
d\Phi(P;k_1,k_2)&=&\frac{1}{16\pi^2} \frac{dx_1dx_2}{x_1x_2}
d^2k_{1\perp}d^2k_{2\perp} \delta (1-x_1-x_2 )
\delta^{(2)}(\vk_{1\perp}+\vk_{2\perp})
\nonumber\\
&\times&\delta \left (s-\frac{m^2+k^2_{1\perp}}{x_1} -
\frac{m^2+k^2_{2\perp}}{x_2} \right )\ ,
\label{2.12}
\eea
where $x_i=k_{iz}/p$,
and the phase space for the triangle diagram reads:
$$
d\Phi(P,P';k_1,k_2,k'_1)=
\frac{1}{16\pi} \frac{dx_1dx_2}{x^2_1x_2}
d^2k_{1\perp}d^2k_{2\perp} \delta (1-x_1-x_2 )
\delta^{(2)}(\vk_{1\perp}+\vk_{2\perp})
$$
\be
\delta \left (s-\frac{m^2+k^2_{1\perp}}{x_1}
-\frac{m^2+k^2_{2\perp}}{x_2} \right )
\delta \left (s'+q^2_\perp-\frac{m^2+(\vk_{1\perp}-\vq_{\perp})
^2}{x_1}
-\frac{m^2+k^2_{2\perp}}{x_2} \right ) .
\label{2.13}
\ee
Then the amplitude $V \to \gamma(q^2) S $ is written as
\be
A_{V \to \gamma(q^2) S } (q^2)=
\frac {Z_{V\to\gamma S}^{(q\bar q)} }{16\pi^3}
\int \limits_{0}^{1}
\frac {dx}{x(1-x)^2}  \int d^2k_{\perp} \psi_{V} (s)
\psi_{S}(s') S_{V \to\gamma  S} (s,s',q^2)\ ,
\label{2.14}
\ee
where $ x=k_{2z}/p$ , $ \vk_{\perp}=  \vk_{2\perp}$, and
the $q\bar q$ invariant masses squared are
\be
s=\frac{m^2+k^2_{\perp} }{x(1-x)}, \qquad
s'=\frac{m^2+(\vk_{\perp}+x \vq_{\perp})^2 }{x(1-x)}\ .
\label{2.15}
\ee

\subsubsection{ Charge factors}
Charge factors for the $n\bar n$ and
$s\bar s$ components in the transition
$\phi\to\gamma f_0$ are determined as:
\be
Z ^{(n\bar n)}_{\phi\to\gamma f_0}=
2\zeta ^{(n\bar n)}_{\phi\to\gamma f_0} =\frac 16\ ,
\label{2.16}
\ee
$$
Z ^{(s\bar s)}_{\phi\to\gamma f_0}=
2\zeta ^{(s\bar s)}_{\phi\to\gamma f_0}=2e_s=-\frac 23\ .
$$
where $\zeta ^{(n\bar n)}_{\phi\to\gamma f_0}$ is the
following convolution:
$\zeta ^{(n\bar n)}_{\phi\to\gamma f_0}=
(u\bar u +d\bar d)/\sqrt{2}\cdot \hat e_q \cdot
(u\bar u +d\bar d)/\sqrt{2}= (e_u+e_d)/2$.
Here $e_u$ and $e_d$ are charges of $u$ and $d$ quarks.
The factor 2 in (\ref{2.16})  is related  to two possibilities for
photon emission, namely, from quark and antiquark.
Likewise, for the process $\phi\to\gamma a_0$, one has:
\be
Z _{\phi\to\gamma a_0}=
2\zeta _{\phi\to\gamma a_0}=
e_u-e_d=1\ .
\label{2.17}
\ee

\subsubsection{ Meson wave functions}

To calculate the form factors, one should define  meson wave functions. The
simplest parameterization is exponential one:
\be
\psi_V(s)=C_V e^{-b_V s}, \qquad
\psi_S(s)=C_S e^{-b_S s}\ .
\label{2.18}
\ee
The parameters $b_V$ and $b_S$ characterize the size of the system,
they are related to the mean radii squared, $R^2_V$ and $R^2_S$,
 of the mesons.
At fixed $R^2_V$ and $R^2_S$ the constants $C_V$ and
$C_S$ are determined by the wave function normalization, which itself
is given by meson form factor in the external field,
$F_{{\rm meson}}(q^2)$, and at small $q^2$ the form factor is:
\be
F_{{\rm meson}}(q^2)\simeq 1+\frac16 R^2_{{\rm meson}}q^2 \ .
\label{2.19}
\ee
The requirement $F_{{\rm meson}}(0)=1$ fixes the constant $C_{{\rm meson}}$
in (\ref{2.18}), while the value $R^2_{{\rm meson}}$ is directly related
to $b_{{\rm meson}}$.

In terms of the light-cone
variables, the form factor $F_{\rm {meson}}(q^2)$
reads:
\be
F_{\rm {meson}} (q^2)=
\frac {1}{16\pi^3}
\int \limits_{0}^{1}
\frac {dx}{x(1-x)^2}  \int d^2k_{\perp} \Psi_{\rm {meson}} (s)
\Psi_{\rm {meson}}(s') S^{\rm (tr)}_{\rm {meson}} (s,s',q^2)\ ,
\label{2.20}
\ee
where $S^{\rm (tr)}_{\rm {meson}}$ is
determined by the following traces:
\begin{eqnarray}
&&2P^\perp_\mu S^{\rm (tr)}_{S}(s,s',q^2)=-{\rm Sp}\left
[(\hat k'_1+m)\gamma^\perp_\mu (\hat k_1+m)(-\hat k_2+m)\right ]\ ,
\\
&&2P^\perp_\mu S^{\rm (tr)}_{V}(s,s',q^2)=- \frac13{\rm Sp}\left
[\gamma'^\perp_\alpha(\hat k'_1+m)\gamma^\perp_\mu (\hat k_1+m)
\gamma^\perp_\alpha(-\hat k_2+m)\right ]  \ .
\nonumber
\label{2.21}
\end{eqnarray}
and the orthogonal components entering (28) are  as follows:
\begin{eqnarray}
&&P^\perp_\mu =P_\mu -q_\mu\frac {(Pq)}{q^2}\ , \qquad
\gamma_\mu^\perp=\gamma_\mu-q_\mu \frac {\hat q}{q^2}\ ,
\nonumber \\
&&\gamma_\alpha^\perp=\gamma_\alpha-P_\alpha \frac {\hat P}{P^2}\ ,
\qquad
 \gamma'^\perp_\alpha=\gamma_\alpha-P'_\alpha \frac {\hat P'}{P'^2}\ ,
\label{2.22}
\end{eqnarray}
where $q=k'_1-k_1$.
When determining $S^{\rm (tr)}_V (s,s',q^2)$, we have averaged over
three polarizations of vector meson that results in the factor $1/3$.

The functions $ S^{\rm (tr)}_{S}(s,s',q^2)$
and  $S^{\rm (tr)}_{V}(s,s',q^2)$ are
equal to:
\begin{eqnarray}
&&S^{\rm (tr)}_{S}(s,s',q^2)=\alpha(s,s',q^2)(s+s'-8m^2-q^2)+q^2\ ,
\\
&&S^{\rm (tr)}_{V}(s,s',q^2)=
\frac23 \left [\alpha(s,s',q^2)(s+s'+4m^2-q^2)+q^2
\right ]\ ,
\nonumber
\label{2.23}
\end{eqnarray}
where $\alpha(s,s',q^2)$ is given in  (\ref{2.8}).

\subsubsection{ Partial width}
The decay partial width is determined as
\be
m_V \Gamma_{V \to \gamma S} =
\frac13 \int d\Phi(p_V;q,p'_S)  |A_{\mu\nu}|^2
\ .
\label{2.24}
\ee
Here the averaging over spin projections of the
$\phi$ meson and summing over photon ones is carried out
(summation over photon spin
variables results in the metric tensor $g_{\mu\mu'}^{\perp\perp}$);
the two-particle phase space
for the radiative decay $V \to \gamma+S$ is
equal to: $\int
d\Phi(p_V;q,p'_S)=(m_V^2-m_S^2)/(16\pi m_V^2)$.
Partial width in terms of the form factor reads:
\be
m_V \Gamma_{V \to \gamma S}=\frac16 \alpha
\frac{m_V^2-m_S^2}{m_V^2} |F_{V \to \gamma S}(0)|^2\; .
\label{2.25}
\ee

\subsection{ The process
$V \to \gamma P $}

The light-cone representation of
the transition
form factor $V \to \gamma_\perp(q^2) P $ reads:
\be
F_{V\to \gamma P} (q^2)=
\frac {Z_{V\to\gamma P} }{16\pi^3}
\int \limits_{0}^{1}
\frac {dx}{x(1-x)^2}  \int d^2k_{\perp} \Psi_V (s) \Psi_{P}
(s')
S_{V\to\gamma P}(s,s',q^2),
\label{2.26}
\ee
The spin factor for pseudoscalar mesons is determined by
\be
-{\rm Sp}\left
[i\gamma_5(\hat k'_1+m)\gamma^{\perp\perp}_\alpha (\hat
k_1+m)\gamma'^\perp_\mu(-\hat k_2+m)\right ]\
=\epsilon_{\mu\alpha\nu_1\nu_2}P_{\nu_1}\tilde q_{\nu_2}
 S_{V\to \gamma P}(s,s',q^2)
\label{2.27}
\ee
where $\tilde q =P-P'$ and $\tilde q_\alpha
\gamma^{\perp\perp}_\alpha=0$, $P'_\mu \gamma'^\perp_\mu =0$. The spin
factor is equal to
\begin{equation} S_{V\to
\gamma P}(s,s',q^2)=4m\, .
\label{2.28}
\end{equation}
Charge factors for the considered radiative decays are as
follows: for the $s\bar s$
component in the reactions $\phi \to \gamma\eta, \gamma\eta'$,
$Z^{(s\bar s)}_{\phi \to \gamma\eta}=Z^{(s\bar s)}_{\phi \to
\gamma\eta'} =-2/3$,
and for the $\pi^0$ and $a_0(980)$ productions,
$ Z_{\phi \to \gamma\pi^0} = Z_{\phi \to \gamma a_0(980)}= 1 $.

Partial width  for the decay $V \to \gamma P $ is equal to
\be
m_V \Gamma_{V \to \gamma P} = \frac13 \int
d\Phi(p_V;q,p'_P) |A_{\mu\nu}|^2
=\frac16 \alpha
\frac{m_V^2-m_P^2}{m_V^2} |F_{V \to \gamma P}(0)|^2\; .
\label{2.29}
\ee

\subsection{Processes $S\to \gamma\gamma$ }

Our calculation of the two-photon decays of scalar and
tensor mesons
is based on the method
developed in \cite{eta} for the study of the pseudoscalar meson
transitions $\pi^0 \to \gamma(q^2)\gamma$, $\eta \to
\gamma(q^2)\gamma$ and $\eta' \to \gamma(q^2)\gamma$,
and for the photon
we use quark-antiquark wave function which was found in
\cite{eta}.
We perform
the calculation of the  scalar and tensor meson transition form
factors $V \to \gamma(q^2)\gamma$ and $T \to
\gamma(q^2)\gamma$ in the region of small $q^2$;  these form
factors, in the limit $q^2 \to 0$, determine partial widths
$S\to \gamma\gamma $ and $T\to \gamma\gamma$.

The transition form factor $q\bar q-meson \to \gamma(q^2)\gamma$ is
determined by the three-point
quark loop diagram of Fig. 1$b$ type that is a convolution
of the $q\bar q$-meson and photon wave functions, $\psi_{q\bar q}
\otimes \psi_{\gamma}$.
Following \cite{eta}, we represent
the photon wave function as a sum of two components which describe the
prompt production of  $q\bar q$ pair at large $s'$ (with a
point-like vertex for the transition $\gamma \to q\bar q $)
as well as at  low-$s'$ region where the
vertex $\gamma \to q\bar q $ has a nontrivial structure due to soft
$q\bar q$ interactions.  The process of Fig. 1$b$ at moderately small
$|q^2|$ is mainly saturated by the contribution of the low-$s'$ region,
in other words, by  soft component of the photon wave function.
The soft component of the photon wave function was restored in
\cite{eta} on the basis of experimental data for the transition
$\pi^0 \to \gamma(q^2)\gamma$ at $|q^2| \leq 1$ GeV$^2$,
it is shown in Fig. 2.

With the
photon wave function found, the decay form factors $S\to
\gamma\gamma$ and $T \to \gamma\gamma$ provide the opportunity
to investigate the scalar and tensor meson wave functions.

\subsubsection{Form factor $S\to \gamma(q^2)\gamma $}

Following the prescription given in
previous sections for $V\to \gamma(q^2) S $ and $V\to \gamma(q^2) P $,
we present the amplitude of the process
$S\to \gamma(q^2)\gamma$ in terms of the
light-cone variables.
The transition form factor $S \to \gamma(q^2) \gamma $ reads:
\be
F_{S\to \gamma\gamma} (q^2,0)=
\frac {Z_{S\to \gamma\gamma} \sqrt {N_c} }{16\pi^3}
\int \limits_{0}^{1}
\frac {dx}{x(1-x)^2}  \int d^2k_{\perp} \psi_{S} (s) \psi_{\gamma}
(s')
S_{S\to \gamma\gamma}(s,s',q^2).
\label{3.1}
\ee
Here, as before in Eq. (\ref{2.14}), the light-cone variables are
introduced as follows: $ x=k_{2z}/p$, $ \vec k_{\perp}= \vec
k_{2\perp}$, and the $q\bar q$ invariant masses squared, $s$ and $s'$,
are determined by Eq. (\ref{2.15}).
The factor $\sqrt {N_c}$, where
$N_c=3$ is the number of colours, is related to the normalization of
the photon wave function performed in  \cite{eta}.

The charge factor $Z_{S\to \gamma\gamma}=2\zeta_{S\to \gamma\gamma}$ is
determined by the quark content of the $S$-meson.
We have two loop diagrams
with quark lines drawn clockwise and anticlockwise: the factor $2$ in
the determination of $Z_{S\to \gamma\gamma}$ stands for this doubling.
For the $f_0$-meson, one has two components with different charge factors
$\zeta_{n\bar n \to
\gamma\gamma}=(e_u^2+e_d^2)/\sqrt 2 $ and $\zeta_{s\bar s\to
\gamma\gamma}=e_s^2 $, while for $a_0$-meson $\zeta_{a_0 \to
\gamma\gamma}=(e_u^2-e_d^2)/\sqrt 2 $.

The spin structure factor $S_{S\to \gamma\gamma}(s,s',q^2)$
is fixed by the three-point quark loop trace
for the amplitude of Fig. 1b, with transverse polarized photons:
\be
-Sp [\gamma^{\perp\perp}_{\beta} (\hat k'_1+m)
\gamma^{\perp\perp}_{\alpha} (\hat k_1+m) (-\hat k_2+m)]
= S_{S\to \gamma\gamma}(s,s',q^2) \;   g_{\alpha\beta}^{\perp\perp}\, .
\label{3.2}
\ee
Here $\gamma^{\perp\perp}_{\beta}$
and $\gamma^{\perp\perp}_{\alpha}$ stand for photon vertices,
$ \gamma^{\perp\perp}_{\alpha} =
g_{\alpha\mu}^{\perp\perp}\gamma_{\mu}$,
while $g_{\alpha\mu}^{\perp\perp}$ is determined by formula (3)
with the following substitution
$q \to P\;-P'$ and $ q' \to P'$.

One has
\be
 S_{S\to \gamma\gamma}(s,s',q^2) = -2m \left [
4m^2-s+s'+q^2-\frac {4ss'q^2}{2(s+s')q^2-(s-s')^2-q^4}
\right ].
\label{3.3}
\ee
Partial width, $\Gamma_{S \to \gamma \gamma}$, is determined as
follows:
\be
m_S\Gamma_{S \to \gamma \gamma} =
\frac 12  \int d\Phi(p_S;q,q')
|A_{\mu\nu} |^2 =
\pi \alpha^2
|F_{S\to \gamma\gamma}(0,0)|^2\, .
\label{3.4}
\ee
Here $ m_S  $ is the  scalar meson mass,
the summation is carried out over outgoing
photon polarizations, the photon identity factor, $1/2$,  is
written explicitly,  and the two-photon invariant phase space is equal
to $d\Phi_2(p_S;q,q') =  1/(16\pi)$.

\subsection{Two-photon tensor meson decay  $T\to \gamma\gamma$}

The decay amplitude $T \to \gamma_\perp (q^2) \gamma  $ can be
considered quite analogousy to the amplitude of the two-photon scalar
meson decay.  The light-cone representation for the form factor
$F^{(H)}_{T\to \gamma\gamma}(q^2,q'^2=0)$ with $H=0,2$ reads:
\be
F^{(H)}_{T\to \gamma\gamma}(q^2,0)=
\frac {Z_{T\to \gamma\gamma} \sqrt {N_c} }{16\pi^3}
\int \limits_{0}^{1}
\frac {dx}{x(1-x)^2}  \int d^2k_{\perp} \psi_T (s) \psi_{\gamma}(s')
S^{(H)}_{T\to \gamma\gamma}(s,s',q^2)
\label{3.5}
\ee
Here we use the same notation as in (\ref{3.1}); the charge factors for
the tensor and scalar mesons are equal to one another,
$Z_{T\to \gamma\gamma}=Z_{S\to \gamma\gamma}$.

The spin structure factors
are fixed by the vertex for transition $T\to q\bar q$;
we denote this vertex as $T_{\mu\nu} $. One has:
\be
S^{(T)}_{\mu\nu,\alpha\beta}= {\rm Sp}
[\gamma^{\perp\perp}_{\beta} (\hat k'_1+m)
\gamma^{\perp\perp}_{\alpha} (\hat k_1+m)T_{\mu\nu} (\hat k_2-m)]
=
\label{3.6}
\ee
$$
S^{(0)}_{\mu\nu,\alpha\beta} (P, \tilde q)
S^{(0)}(P\; ^2,P\; '^2,\tilde q^2)+
S^{(2)}_{\mu\nu,\alpha\beta} (P,\tilde q)
S^{(2)}(P\; ^2,P\; '^2,\tilde q^2),
$$
where $\gamma^{\perp\perp}_{\alpha}$
and $\gamma^{\perp\perp}_{\beta}$ stand for photon vertices,
$ \gamma^{\perp\perp}_{\alpha} =
g_{\alpha\alpha '}^{\perp\perp}\gamma_{\alpha '}$,
and $g_{\alpha\alpha '}^{\perp\perp}$ is determined by Eq. (3)
with the following substitution
$q \to P-P'$ and $ q' \to P'$. The moment-operators
$S^{(0)}_{\mu\nu,\alpha\beta} (P,\tilde q)$ and
$S^{(2)}_{\mu\nu,\alpha\beta} (P,\tilde q)$ work also in the
intermediate state momentum space.
Recall that $\tilde q =P-P'$ and $P^2=s$, $P'^2=s'$, $\tilde q^2= q^2$,
while the momenta $k'_1$, $k_1$ and $k_2$ are mass-on-shell.

The vertex $T_{\mu\nu}(k) $ in its minimal form reads:
\be
T_{\mu\nu}(k)  = k_\mu \gamma_\nu + k_\nu \gamma_\mu -\frac 23
g_{\mu\nu}^\perp \hat k\ ,
\label{3.7}
\ee
where $k=k_1-k_2$ and $g_{\mu\nu}^\perp P_\nu =0$.

Spin structure factors $S^{(0)} (s,s', q^2)$
and $S^{(2)}(s,s', q^2)$ are calculated by  projecting (\ref{3.6})
on the moment-operators
$ S^{(H)}_{\mu\nu,\alpha\beta} (P, \tilde q)$:
\be
S^{(H)}(s,s', q^2)=
\frac{ S^{(H)}_{\mu\nu,\alpha\beta} (P, \tilde q)
S^{(T)}_{\mu\nu,\alpha\beta} }
{\left (S^{(H)}_{\mu'\nu',\alpha'\beta'} (P, \tilde q) \right )^2}
\label{3.8}
\ee
Explicit expressions for
the spin structure factors $S^{(0)} (s,s', q^2)$
and $S^{(2)}(s,s', q^2)$ are rather cumbersome, and we do not present
them here.

The $q\bar q\; (2^{++})$  state can be constructed in two ways, namely,
with the $q\bar q$ orbital momenta $L=1$ and $L=3$
(the $^3P_2q\bar q$ and $^3F_2q\bar q$ states). The vertex
$T_{\mu\nu}$ of Eq. (\ref{3.7}), corresponding to the dominant
$P$-wave $q\bar q$ state, includes also certain admixture of the
$F$-wave $q\bar q$ state.

The vertex
for the production of pure $q\bar q\; (L=1)$ state reads:
\be
T^{(L=1)}_{\mu\nu} = k_\mu \Gamma_\nu + k_\nu \Gamma_\mu -\frac 23
g_{\mu\nu}^\perp (\Gamma k), \qquad
\Gamma_\mu =\gamma ^\perp _\mu -\frac{k_\mu}{2m+\sqrt s}\; ,
\label{3.7a}
\ee
where the operator $\Gamma_\mu $ selects the spin-1 state for the
$q\bar q$ (see \cite{AKMS,moment,spin} for detail).

The $(L=3)$-operator for the $^3F_2q\bar q$ state is equal to:
\be
T^{(L=3)}_{\mu\nu} = k_\mu k_\nu (\Gamma k) -\frac {k^2}{5}
\left (
g_{\mu\nu}^\perp (\Gamma k) +
\Gamma_\mu k_\nu + \Gamma_\nu k_\mu \right ).
\label{3.7b}
\ee
For the $q\bar q$ wave function
of  tensor mesons, we use a parametrization similar to that
for scalar mesons, see Eq. (\ref{2.18}). The parameters $C_T$ and
$b_T$ are
determined by the tensor meson charge form factor
at small $q^2$, Eq. (\ref{2.19}): the charge form factor
is given by Eq. (\ref{2.20}).
The spin factor for the tensor meson,
$S_T(s,s',q^2)$ is defined by the quark
loop trace as follows:
\be
\frac 15 {\rm Sp} [T_{\mu\nu} (\hat k_1+m)
\gamma_{\alpha} (\hat k'_1+m)T'_{\mu\nu}  (\hat k_2-m)] =
2P_{\perp \alpha} S_T(s,s',q^2),
\label{3.9}
\ee
The operator $T_{\mu\nu}$ is written for the initial state transition
$T\to q\bar q$, Eq. (\ref{3.7}), while $T'_{\mu\nu}$ describes the
production of the outgoing tensor meson $ q\bar q\to T$ that
requires the following
substitutions in  (\ref{3.7}): $k_1\to k'_1$ and $P\to P'$.  The tensor
meson charge form factor is averaged over polarizations that
results in the factor $1/5$ in (\ref{3.9}).

Partial width, $\Gamma_{T \to \gamma \gamma}$, is determined as
\bea
m_{T}\Gamma_{T \to \gamma \gamma}& =&
\frac 12  \int d\Phi(p_T;q,q')
\frac 15
\sum_{\mu\nu,\alpha\beta} |A_{\mu\nu,\alpha\beta}|^2\\
\nonumber
& =& \frac 45
\pi \alpha^2\left [ \frac 13
| F^{(0)}_{T\to \gamma\gamma}(0,0)|^2 +
| F^{(2)}_{T\to \gamma\gamma}(0,0)|^2 \right ].
\label{3.10}
\eea
Here $m_T$ is the tensor meson mass,
the summation is carried out over outgoing
photon polarizations, the photon identity factor, $ 1/2$,  is written
explicitly; the averaging over  tensor-meson polarizations results in
the  factor $1/5$.

\section{Results}
In this Section we present the results of calculations of
partial widths for the radiative decays
$\phi(1020)\to\gamma f_0(980)$,
$\phi(1020)\to\gamma\eta, \gamma\eta', \gamma\pi^0,
\gamma a_0(980)$ and two-photon decays
$f_0(980)\to \gamma\gamma$,
$a_0(980)\to \gamma\gamma$,  $a_2(1320)\to \gamma\gamma$,
$f_2(1270)\to \gamma\gamma$, $f_2(1525)\to \gamma\gamma$.

\subsection{$\phi(1020)\to\gamma\lowercase{f}_0(980)$:
the decay amplitude and  partial width}

Here
we calculate the branching ratio for the decay
$\phi(1020) \to \gamma f_0(980)$ assuming
the $q\bar q$ structure of  $ f_0(980)$.

\subsubsection{ Wave functions of  $\phi(1020)$ and $
\lowercase{f}_0(980)$}
We write the wave functions of $\phi(1020)$ and $f_0(980)$
as follows:
\begin{eqnarray}
&&\Psi_\phi(s)=\left (n\bar n\sin \varphi_V  + s\bar s\cos \varphi_V
\right ) \psi_\phi(s)\ ,  \\
&&\Psi_{f_0(980)}(s)=\left ( n\bar n\cos \varphi+ s\bar s
\sin \varphi\right ) \psi_{f_0(980)}(s) \ ,
\nonumber
\label{4.1}
\end{eqnarray}
assuming similar $s$-dependence for the $n\bar n$ and $s\bar
s$ components.
For $\psi_\phi(s)$ and
$\psi_{f_0(980)}(s)$ the exponential parameterization is
used, Eq. (\ref{2.18}).
The radius
squared of $n\bar n$ component in the $\phi$-meson is suggested to
be approximately the same as that of the pion:
$R^{2}_\phi (n\bar n)\simeq 10.9$ GeV$^{-2}$, while the radius
squared for the $s\bar s$ component,
$R^{2} _\phi (s\bar s)$,
appears to be slightly less, $R^{2}
_\phi (s\bar s) \simeq 9.3$ GeV$^{-2}$, that corresponds to
$b_\phi = 2.5$ GeV$^{-2}$.
As to $f_0(980)$, we vary the radius of $n\bar n$ component
in the interval $6\;{\rm GeV}^{-2} \le R^2 _{f_0(980)}(n\bar n) \le
18\;{\rm GeV}^{-2}$.

\subsubsection{Partial width $\phi(1020)\to\gamma
\lowercase{f}_0(980)$}
The amplitude $A_{\phi \to \gamma f_0(980)} (0)$
is the sum of two terms related to the
$n\bar n$ and $s\bar s$ components:
\begin{equation}
A_{\phi \to \gamma f_0} (0)
=\cos \varphi\sin \varphi_V F^{(n\bar n)}_{\phi \to \gamma
f_0(980)} (0)+
\sin \varphi\cos \varphi_V F^{(s\bar s)}_{\phi \to \gamma
f_0(980)} (0)\ .
\label{4.2}
\end{equation}
In our estimations we put $\cos \varphi_V \sim 0.99$ and,
correspondingly, $|\sin \varphi_V| \sim 0.14$;
for the $f_0(980)$ we vary the mixing angle in the interval
$0^\circ \le |\varphi|\le 90^\circ$.

The results of the calculation are shown in Figs. 3 and 4. In Fig. 3$a$
the values $A^{(n\bar n)}_{\phi \to \gamma f_0(980)} (0)$
and $A^{(s\bar s)}_{\phi \to \gamma f_0(980)} (0)$
are plotted versus radius
squared, $R^2_{f_0(980)}$.

In Fig. 4 one can see the value ${\rm BR}(\phi\to\gamma f_0(980))$ at
various $\varphi $. Shaded areas
correspond to the variation of $\varphi_V$ in the interval $-8^\circ
\le \varphi_V \le 8^\circ$; the lower and upper curves of the shaded
area correspond to the destructive and constructive interferences of
$A^{(n\bar n)}_{\phi \to \gamma f_0(980)} (0)$ and
$A^{(s\bar s)}_{\phi \to \gamma f_0(980)} (0)$,
respectively.

The measurement of the $f_0(980)$ signal in the $\gamma \pi^0\pi^0$
reaction (SND Collaboration) gives the branching ratio
${\rm BR}(\phi\to\gamma f_0(980))=(3.5\pm 0.3 ^{+1.3}_{-0.5})\times 10^{-4}$
\cite{novosib}; in the analysis of $\gamma \pi^0\pi^0$ and
$\gamma \pi^+\pi^-$ channels (CMD Collaboration) it was found
${\rm BR}(\phi\to\gamma f_0(980))=
(2.90\pm 0.21 \pm 1.5)\times  10^{-4}$ \cite{novosib}; the averaged
value is given in \cite{PDG-00}:
${\rm BR}(\phi\to\gamma f_0(980))=(3.4\pm 0.4)\times 10^{-4}$.
In our
estimation of the permissible interval for the mixing angle $\varphi$,
we have used the averaged value given by \cite{PDG-00}, with the
inclusion of systematic errors of the order of those found in
\cite{novosib}:  ${\rm BR}(\phi\to\gamma f_0(980))=(3.4\pm 0.4
^{+1.5}_{-0.5})\times 10^{-4}$.

The calculated values of ${\rm BR}(\phi\to\gamma f_0(980))$
agree with experimental data for
$|\varphi|\ge 25^\circ$; larger values of  mixing angle, $|\varphi|$
$\geq 55^\circ$,
correspond to a more compact structure of $f_0(980)$, namely,
$R^2_{f_0(980)} \leq 10$ GeV$^{-2}$, while small mixing
angles $|\varphi|\sim 25^\circ$ are related to a loosely bound
structure of the $f_0(980)$, $R^2_{f_0(980)} \geq 12$ GeV$^{-2}$.

The evaluation of the radius
of $f_0(980)$ was performed in \cite{radius} on the basis of GAMS data
\cite{GAMS}, where the $t$-dependence was measured in the process
$\pi^-p \to f_0(980)\,n$ ($t$ is the momentum squared transferred to
$f_0(980)$):  these data favour a comparatively compact structure of
the $q\bar q$ component in $f_0(980)$, namely, $ R^2_{f_0(980)} =6\pm
6$ GeV$^{-2}$.

\subsection{Radiative decays
$\phi(1020)\to\gamma\eta, \gamma\eta', \gamma\pi^0,
\gamma\lowercase{a}_0(980)$}

 The decays $\phi(1020) \to \gamma\eta, \gamma\eta'$ do
not provide us with  a direct information on the quark content of
$f_0(980)$ and $\phi (1020)$; still, calculations and
comparison with data
are necessary to check the reliability of the
method.  The decays $\phi(1020) \to \gamma\pi^0$
and $\phi(1020) \to \gamma a_0(980)$ allow us to evaluate
the admixture of the $n\bar n$ component in the $\phi$ meson; as
is seen in the previous section, this admixture affects
significantly the value $\Gamma_{\phi(1020) \to\gamma f_0(980)}$.

For the transitions $\phi \to \gamma\eta$ and $\phi \to \gamma\eta'$
we take into account the dominant $s\bar s$ component only:
$-\sin\theta \, s\bar s$ in $\eta$-meson and
$\cos\theta \, s\bar s$ in $\eta'$-meson, with $\sin\theta =0.6$.

For the pion wave function we have chosen $b_\pi=2.0$ GeV$^{-2}$ that
corresponds to $R^2_\pi=10.1$ GeV$^{-2}$, the same radius is fixed
for the $n\bar n$ component in $\eta$ and $\eta'$.
As to the strange component in $\eta$ and $\eta'$, we put its slope to
be the same: $b_{\eta(s\bar s)}=b_{\eta'(s\bar s)}=2$ GeV$^{-2}$, that
leads to a smaller radius $R^2(s\bar s)=8.3$ GeV$^{-2}$.

The calculation results for branching ratios compared to those
given by
PDG-com\-pi\-la\-tion \cite{PDG-00} are as follows:
\begin{eqnarray}
&&{\rm BR}(\phi\to \eta\gamma)=1.46\times 10^{-2}\ , \;\;
{\rm BR}_{{\rm PDG}}(\phi\to \eta\gamma)=(1.30\pm 0.03)\times 10^{-2}\ ,
\nonumber
 \\
&&{\rm BR}(\phi\to \eta'\gamma)=0.97\times 10^{-4}\ , \;\;
{\rm BR}_{{\rm PDG}}(\phi\to \eta'\gamma)=(0.67 ^{+ 0.35}_{-0.31})\times
10^{-4}.
\end{eqnarray}
As is clearly seen, the calculated branching ratios agree
reasonably with those given in \cite{PDG-00}.

For the process $\phi \to \gamma \pi^0$ the compilation \cite{PDG-00}
gives ${\rm BR}(\phi\to\gamma \pi^0 )=(1.26\pm \pm 0.10)\times
10^{-3}$, and this value requires $|\sin\varphi_V|\simeq 0.07$ (or
$|\varphi_V|\simeq 4^\circ$), for just with this admixture of the
$n\bar n$ component in $\phi (1020)$ we reach the agreement with data.
However, in the estimation of the allowed regions for  mixing angle
$\varphi$, Fig. 4, we use
\begin{equation}
|\varphi_V| = 4^\circ\pm 4^\circ
\end{equation}
considering the accuracy inherent to the quark model to be
comparable with the obtained small value of $|\varphi_V|$.

The process
$\phi(1020) \to \gamma a_0(980)$ depends also on the mixing angle
$|\varphi_V|$: the decay amplitude is proportional to $\sin \varphi_V$,
namely, $A_{\phi \to \gamma a_0} =
\sin \varphi_V A^{(n\bar n)}_{\phi \to \gamma a_0} $.
For the region $R^2_{a_0(980)} \sim 8$ GeV$^{-2}$ -- $12$
GeV$^{-2}$, our calculation gives the following branching ratio:
\begin{equation}
{\rm BR}\left (\phi (1020) \to \gamma a_0 (980)
\right )= \sin^2 \varphi_V \cdot (14\pm 3) \times 10^{-4}
\label{phi-a0}
\end{equation}
with lower values for $R^2_{a_0(980)} \sim 8$ GeV$^{-2}$ and larger ones
for $R^2_{a_0(980)} \sim 12$ GeV$^{-2}$. At $\sin^2 \varphi_V =
0.01\pm 0.01$, we have ${\rm BR}(\phi (1020) \to \gamma a_0 (980) )=
(0.14\pm 0.14)  \times10^{-4}$.

In \cite{novosib}, the $\eta\pi^0$ spectrum was measured
in the radiative decay
$\phi (1020) \to \gamma \eta\pi^0$: it was found that
${\rm BR}(\phi (1020) \to \gamma \eta\pi^0;\; M_{\eta\pi} > 900$ MeV$)=
(0.46\pm0.13)\times 10^{-4}$. This value does not contradict the
equation (\ref{phi-a0}) with $|\varphi_V|=8^\circ$; moreover, if the
ratio of the background/resonance in the region $ M_{\eta\pi} \sim 900$
MeV is not small, that is rather possible, the value found in
\cite{novosib}  agrees with smaller values of $|\varphi_V| $.

\subsection{Radiative decays $\lowercase{f}_0(980)\to \gamma \gamma $
and $\lowercase{a}_0(980)\to \gamma \gamma $}

The amplitude $A_{f_0(980) \to \gamma \gamma} (0,0)$ is determined
by contributions of two flavour components:
\begin{equation}
A_{f_0(980) \to \gamma \gamma} (0,0)=
\cos \varphi F^{(n\bar n)}_{f_0 (980)\to \gamma \gamma} (0,0) +
\sin \varphi F^{(s\bar s)}_{f_0 (980)\to \gamma \gamma} (0,0)\; .
\label{6.3}
\end{equation}
The amplitudes $A^{n\bar n}_{f_0 (980)\to \gamma \gamma} (0,0)$ and
$A^{s\bar s}_{f_0 (980)\to \gamma \gamma} (0,0)$ depend
on the radius squared of $f_0 (980)$: these amplitudes plotted versus
$R^2_{f_0(980)}$ are shown in Fig. 3$b$.

Figure 5 demonstrates the comparison of calculated
partial width
$\Gamma_{f_0(980) \to \gamma \gamma}$, at different $R^2_{f_0(980)}$ and
$\varphi$, with the magnitude found in \cite{pennington}:
$\Gamma_{f_0(980)\to\gamma\gamma}=0.28^{+0.09}_{-0.13}\; $.
It is possible to describe the data using positive
mixing angles $77^\circ \le \varphi \le 93^\circ$  as well as negative
ones: $(-54^\circ) \le \varphi \le (-38^\circ)$.

The amplitude for the decay $a_0(980)\to\gamma\gamma$ is
determined by the similar form factor that is similar
to that of $n\bar n$
component in the $f_0(980)$, with the only difference
$\zeta_{f_0}\to \zeta_{a_0} $.
The amplitude
$A_{a_0(980)\to\gamma\gamma}/2\zeta_{a_0(980)\to\gamma\gamma}$
is shown in Fig. 3$b$
as a function of $R^2_{a_0(980)}$.  Experimental study of $\Gamma
(a_0(980)\to\gamma\gamma)$ was carried out in Refs. \cite{a0f0,a0}, the
averaged value is: $\Gamma (\eta \pi)
\Gamma (\gamma\gamma)/\Gamma_{total}=0.24^{+0.08}_{-0.07}$ keV
\cite{PDG-00}.  Using $\Gamma_{total} \simeq \Gamma (\eta \pi)+\Gamma
(K \bar K)$, we have $\Gamma
(a_0(980)\to\gamma\gamma)=0.30^{+0.11}_{-0.10}$ keV.  The calculated
value of $\Gamma (a_0(980)\to\gamma\gamma)$ agrees with data at
$R^2_{a_0(980)}$ belonging to the interval
$10$ GeV$^{-2}\le R^2_{a_0(980)}\le 26$ GeV$^{-2} $: the
values of $R^2_{a_0(980)}$ of the order of $\sim 10$ -- $17$ GeV$^{-2}
$ look quite reasonable for a meson of the $1^3P_0 q\bar q$ multiplet.

\subsection{Partial widths for the two-photon decays of tensor
mesons}

Here we present our results for the decays
$a_2(1320)\to \gamma\gamma$, $f_2(1270)\to \gamma\gamma$ and
$f_2(1525)\to \gamma\gamma$.

\subsubsection{Decay $\lowercase{a}_2(1320)\to \gamma\gamma$}

The form factor $F^{(H)}_{a_{2}(1320)\to \gamma\gamma}(0,0)$ is equal
to that for the $n\bar n$ component, up to the charge factor:
\be
F^{(H)}_{a_{2}(1320)\to \gamma\gamma}(0,0)/Z_{a_{2}(1320)}=
F^{(H)}_{n\bar n\to \gamma\gamma}(0,0)/Z_{n\bar n} \ ,
\ee
This universal form factor is determined by
the spin structure of the vertex $a_2(1320) \to q\bar q$, which
is regulated by the admixture of the $^3F_2 q\bar q$-state in
$a_2(1320)$. Figure 6
demonstrates the calculated form factor $F^{(H)}_{n\bar n\to
\gamma\gamma}(0,0)/Z_{n\bar n}$ as a functions of $R^2_T$ for different
vertices given by  Eqs. (\ref{3.7}), (\ref{3.7a}) and (\ref{3.7b}).

The comparison with data is performed for the
form factors calculated with the  use of minimal vertex given by Eq.
(\ref{3.7}). In Fig. 7 we plot calculated values of
$\Gamma_{a_2(1320) \to \gamma\gamma}$ being a  function of
$R^2_{a_2(1320)}$
versus the data: $\Gamma_{a_2(1320) \to \gamma\gamma}=0.98\pm0.05\pm
0.09$ keV \cite{l3} and $\Gamma_{a_2(1320) \to
\gamma\gamma}=0.96\pm0.03\pm 0.13$ keV \cite{argus}.
The calculated value of $\Gamma_{a_2(1320) \to \gamma\gamma}$
reproduces data with $R^2_{a_2(1320)}
\la 9$ GeV$^{-2}$ only.
Still, one should not
be convinced that  larger values of the radius
$R^2_{a_2(1320)}$ are excluded by the data.
Experimental extraction of the signal $a_2(1320) \to \gamma\gamma$
faces the problem of a correct account for coherent background.
This problem has been investigated in \cite{argus90}:
it was shown
that the measured value of  $\Gamma_{a_2(1320) \to \gamma\gamma}$ can
fall down by a factor $\sim 1.5$ due to the interference
"signal--background". Therefore, being careful, we estimate
the region for  $\Gamma_{a_2(1320) \to \gamma\gamma}$ allowed by
the data as
$1.12$ keV$\leq \Gamma_{a_2(1320) \to \gamma\gamma}\leq 0.60$ keV.

Comparing the calculations with data, one should take into account
uncertainties inherent to the model. In our calculation,
$\Gamma_{a_2(1320) \to \gamma\gamma}$ strongly depends on
the constituent quark mass.
In Fig. 6, the form factors and partial widths are depicted
for $m_{u,d}=350$ MeV and $m_s=500$ MeV. However,  decreasing of
constituent quark mass by 10\%
results in the increase of the form factor $F^{(s)}_{n\bar
n}(0,0)$  approximately by 10\%, that means the 20\% growth
of the calculated value of $\Gamma_{a_2(1320) \to \gamma\gamma}$ at
fixed $R^2_{a_2(1320)}$. The 10\% uncertainty in the definition of the
constituent quark mass looks quite reasonable, therefore,
the 20\% error in the model prediction for
$\Gamma_{a_2(1320) \to \gamma\gamma}$ is to be regarded as
normal.

Summing up, the calculation of $\Gamma_{a_2(1320) \to \gamma\gamma}$
with the minimal vertex $a_2(1320) \to q\bar q$ given by
(\ref{3.7}) provides  reasonable agreement  with data at $7$
GeV$^{-2}\la R^2_{a_2(1320)} \la 13$ GeV$^{-2}$.

The vertex corresponding to the production of
$q\bar q$ pair in the $F$-wave, Eq. (\ref{3.7b}), gives the partial
width value $\sim 0.1$ keV that is by an order of value less than for
the $P$-wave $q\bar q$ component.

The vertex for pure $P$-wave $q\bar q$ state, see (\ref{3.7a}),
gives us a 10\% smaller value of $\Gamma_{a_2(1320) \to
\gamma\gamma}$ as compared to what is provided by minimal vertex
(\ref{3.7}).

\subsubsection{The decays $\lowercase{f}_2(1270) \to \gamma\gamma$ and
$\lowercase{f}_2(1525) \to \gamma\gamma$ }

We define
the wave functions of $f_2(1270)$ and $f_2(1525)$ as follows:
\bea
&&\Psi_{f_2(1270)}(s)=\left ( \cos\varphi_T\; n\bar n +
\sin \varphi _T\; s\bar s \right ) \psi_T (s) ,
\nonumber \\
&&\Psi_{f_2(1525)}(s)= \left (
-\sin\varphi_T\; n\bar n + \cos \varphi _T \; s\bar s
\right ) \psi_T (s) .
\eea
Then, the form factors for the two-photon decays
of $f_2$-mesons read:
\bea
F^{(H)}_{f_2(1270)\to \gamma\gamma}(0,0)&=&  \cos\varphi_T
F^{(H)}_{n\bar n\to \gamma\gamma}(0,0)
 + \sin \varphi _T
F^{(H)}_{s\bar s\to \gamma\gamma}(0,0) , \nonumber \\
F^{(H)}_{f_2(1525)\to \gamma\gamma}(0,0)&=&
-\sin\varphi_T
F^{(H)}_{n\bar n\to \gamma\gamma}(0,0)
 + \cos \varphi _T
F^{(H)}_{s\bar s\to \gamma\gamma}(0,0) .
\eea
Following \cite{PDG-00,l3,argus,argus90},
we put the following values for partial
widths: $\Gamma_{f_2(1270) \to \gamma\gamma}=
(2.60\pm{0.25}^{+0.00}_{-0.25})$ keV and
$\Gamma_{f_2(1525) \to \gamma\gamma}=
(0.097\pm{0.015}^{+0.00}_{-0.25})$ keV. The magnitude of the extracted
signal depends on the type of  model used for the description of the
background. With coherent background, the magnitude of the signal
decreases, and the second error in $\Gamma_{f_2(1270) \to
\gamma\gamma}$ and $\Gamma_{f_2(1525) \to \gamma\gamma}$ is related to
the background uncertainties.

A satisfactory description of data has been reached with $R^2_{T} \la
10$ GeV$^{-2}$ and two
mixing angles $\varphi_T$: $\varphi_T \simeq 0^\circ$
and $\varphi_T \simeq 25^\circ$, see Fig. 8. For example, at
$R^2_{f_T}=9$ GeV$^{-2}$ and $\varphi_T = 0^\circ$, the
calculations result in $\Gamma_{f_2(1270) \to \gamma\gamma}=2.240$ keV
and $\Gamma_{f_2(1525) \to \gamma\gamma}=0.090$ keV.  Nearly the
same values are reproduced at
$R^2_{f_T}=9$ GeV$^{-2}$ and $\varphi_T = 25^\circ$, namely,
$\Gamma_{f_2(1270) \to \gamma\gamma}=2.237$ keV
and $\Gamma_{f_2(1525) \to \gamma\gamma}=0.093$ keV.

As for the reaction $a_2(1320) \to \gamma\gamma$, the model
uncertainties,  $\sim 20\%$, are inherent in calculations of
$\Gamma_{f_2(1270) \to \gamma\gamma}$
and $\Gamma_{f_2(1525) \to \gamma\gamma}$.

\section{Conclusion}

Figure 9 demonstrates the $(\varphi , R^2_{f_0(980)})$-plot
where the allowed areas for the reactions $\phi (1020) \to \gamma f_0
(980)$ and $f_0(980) \to \gamma \gamma$ are shown. We see that
radiative decays
$\phi(1020) \to\gamma f_0(980)$ and $f_0(980) \to \gamma\gamma$
are well described in the framework of the hypothesis of the
dominant $q\bar q$
structure of $f_0(980)$. The solution with negative $\varphi$ seems
more preferable. For this solution
the mixing angle $\varphi$ for $n\bar n$
and $s\bar s$ components ($n\bar n\cos \varphi +s\bar s\sin \varphi$)
is equal to $\varphi=- 48^\circ \pm 6^\circ$, that is, the $q\bar q$
component is rather close to the flavour octet
($\varphi_{{\rm octet}}=-54.7^\circ$). However, the radiative-decay
data do not exclude the variant when the $f_0(980)$ is almost pure
$s\bar s$ state with $\varphi=85^\circ\pm 5^\circ$.

The dominance of the quark-antiquark component does not exclude the
existence of other components in $f_0(980)$
on the level $10\% -20\% $. The location of
resonance pole near the $K\bar K$ threshold definitely points to
certain admixture of the long-range $K\bar K$ component in $f_0(980)$.
To investigate this admixture the precise measurements of the $K\bar
K$ spectra in the interval 1000---1150 MeV are necessary: only these
spectra could shed the light on the role of the long-range $K\bar K$
component in $f_0(980)$.

The existence of the long-range $K\bar K$ component or that of gluonium
in the $f_0(980)$ results in a decrease of the $s\bar s$ fraction in
the $q\bar q$ component: for example, if the long-range $K\bar K$ (or
gluonium) admixture is of the order of 15\%, the data require either
$\varphi=- 45^\circ \pm 6^\circ$ or $\varphi=83^\circ \pm 4^\circ$.

There is no problem with description of the decay $a_0(980)\to
\gamma\gamma$ within the hypothesis about the $q\bar q$ origin of
the  $a_0(980)$: the data are in a good agreement with the results
of calculations at $R^2_{a_0(980)} \sim 10$ -- $17$ GeV$^{-2} $.

We have calculated the two-photon decays of tensor mesons, members of
the $q\bar q$ multiplet $1^3P_2q\bar q$.
The calculated partial widths of radiative decays,
$a_2(1320)\to \gamma\gamma$, $f_2(1270)\to \gamma\gamma$ and
$f_2(1525)\to \gamma\gamma$, are in a reasonable agreement with
data. The radial wave functions of $a_2(1320)$, $f_2(1270)$ and
$f_2(1525)$ are close to those of $a_0(980)$ and
$f_0(980)$ found in the study of radiative decays
$a_0(980)\to \gamma\gamma$, $f_0(980)\to \gamma\gamma$ and
$\phi(1020)\to \gamma f_0(980)$. The possibility to describe
simultaneously scalar and tensor mesons using approximately equal
radial wave
functions may be considered as a strong argument that all these mesons,
tensor ones, $a_2(1320)$, $f_2(1270)$ and $f_2(1525)$ and scalar ones,
$a_0(980)$ and $f_0(980)$, are members of the same $P$-wave $q\bar
q$ multiplet.

We are grateful to L.G. Dakhno,
V.N. Markov, M.A. Matveev and A.V. Sarantsev for
useful discussions. The paper is suppored by RFBR grant 01-02-17861.

\newpage

\begin{figure}
\centerline{\epsfig{file=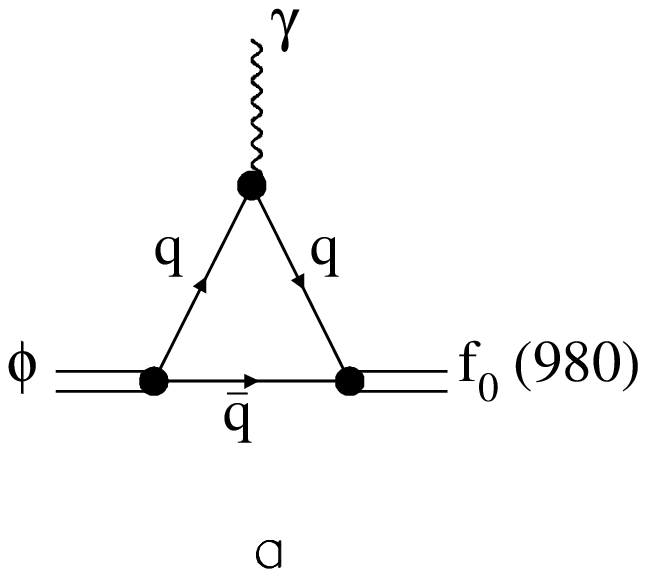,height=6cm}\hspace{1cm}
            \epsfig{file=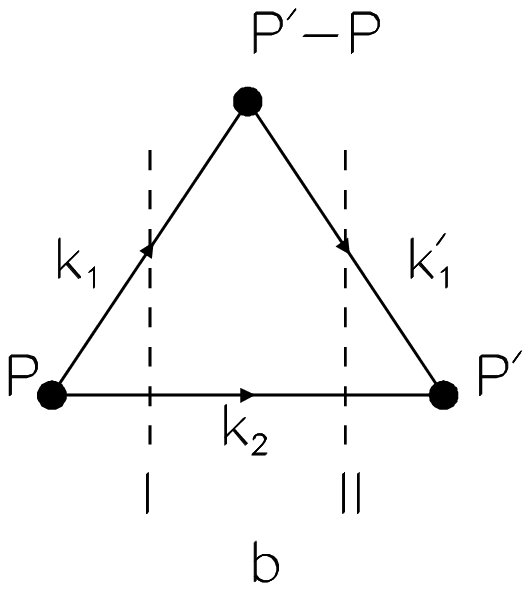,height=6cm}}
\caption{a) Diagrammatic representation of the transition
$\phi(1020)\to\gamma f_0(980)$. b) Three-point quark diagram: dashed
lines I and II mark two cuttings in the double spectral
representation. }
\end{figure}

\begin{figure}
\centerline{\epsfig{file=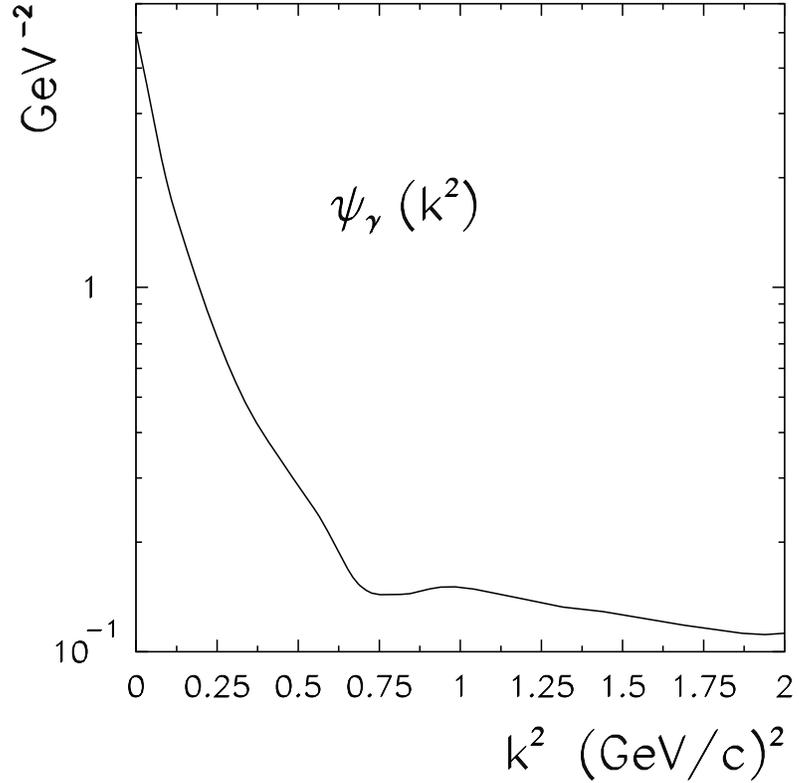,width=8cm}}
\caption{Photon wave function for non-strange quarks,
$\psi_{\gamma\to n\bar n}(k^2)=g_\gamma(k^2)/(k^2+m^2)$, where
$k^2=s/4-m^2$; the wave function for the $s\bar s$
component is equal to
$\psi_{\gamma\to s\bar s}(k^2)=g_\gamma(k^2)/(k^2+m^2_s)$; the
constituent quark masses are  $m$=350 MeV and $m_s$=500 MeV.}
\end{figure}

\begin{figure}
\centerline{\epsfig{file=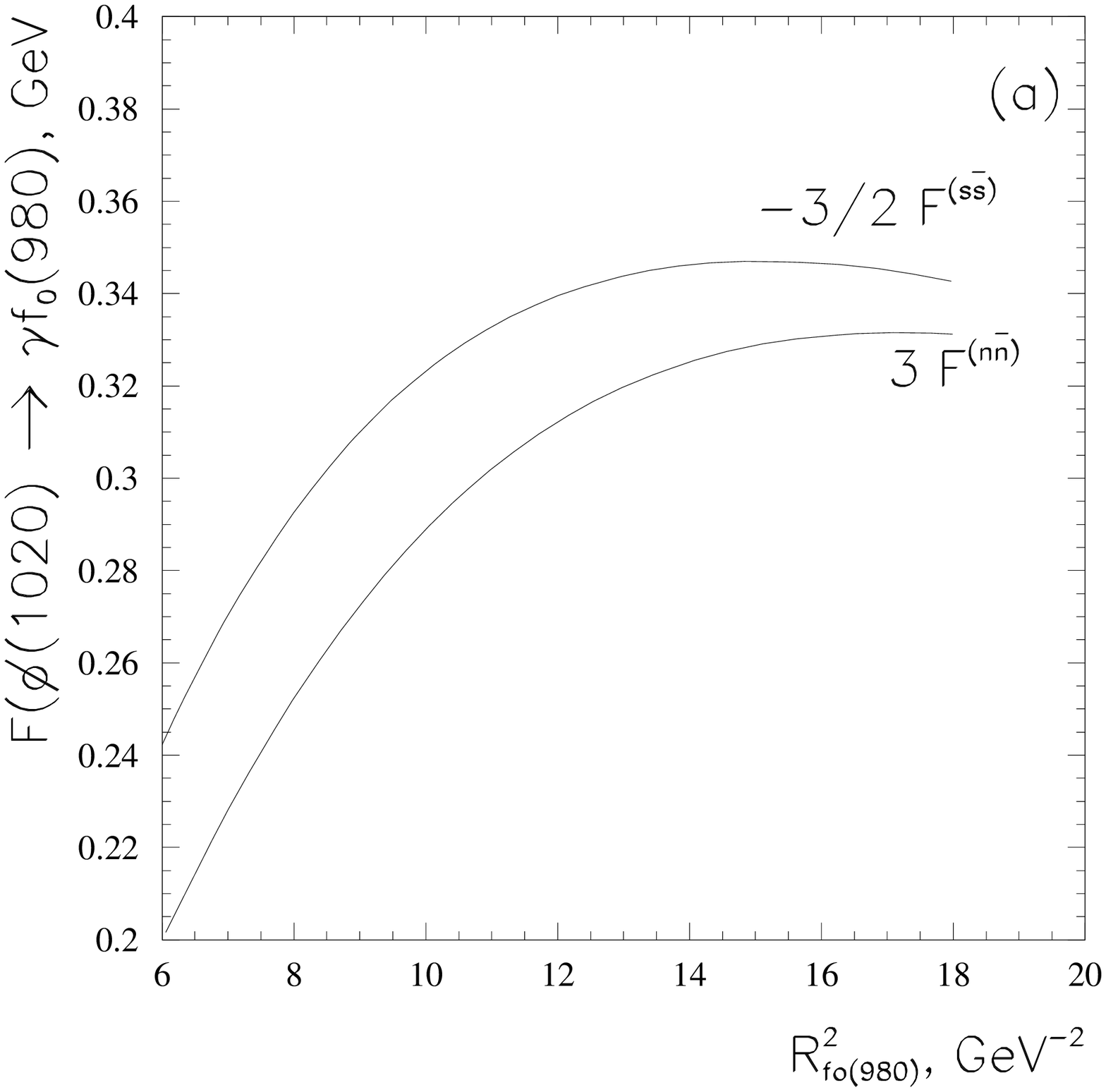,width=8cm}\hspace{0.5cm}
            \epsfig{file=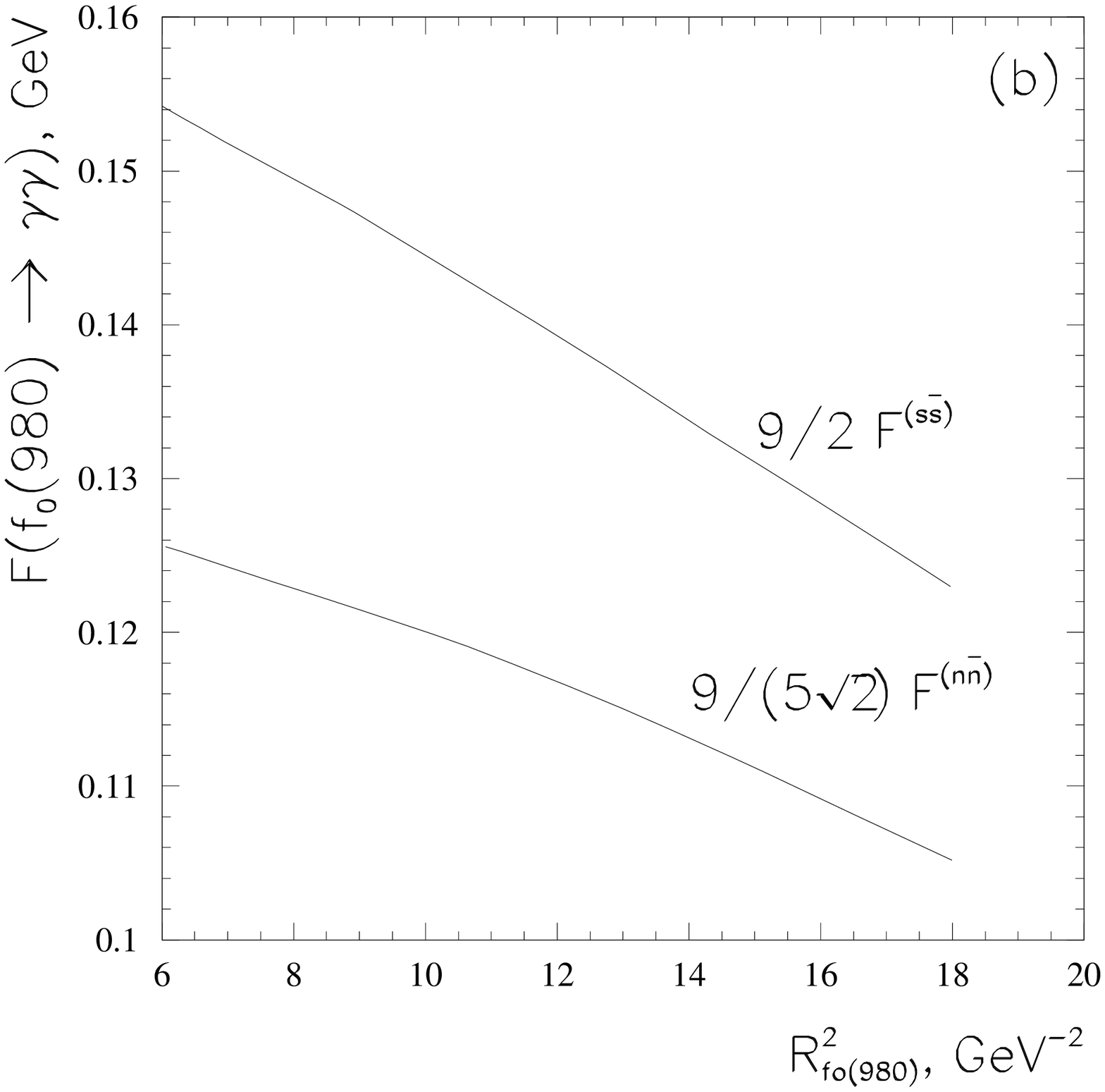,width=8cm}}
\caption{Amplitudes for strange and non-strange components,
$s\bar s$ and $n\bar n$, as functions of the
$f_0(980)$-meson radius squared: a)
$F^{(n\bar n)}_{\phi\to\gamma f_0}(0)/
Z^{(n\bar n)}_{\phi\to\gamma f_0}$ and
$F^{(s\bar s)}_{\phi\to\gamma f_0}(0)/
Z^{(s\bar s)}_{\phi\to\gamma f_0}$,
b) $F^{(n\bar n)}_{f_0\to\gamma\gamma}(0)/
Z^{(n\bar n)}_{f_0\to\gamma\gamma}$
and $F^{(s\bar s)}_{f_0\to\gamma\gamma}(0)/
Z^{(s\bar s)}_{f_0\to\gamma\gamma}$.}
\end{figure}

\begin{figure}
\centerline{\epsfig{file=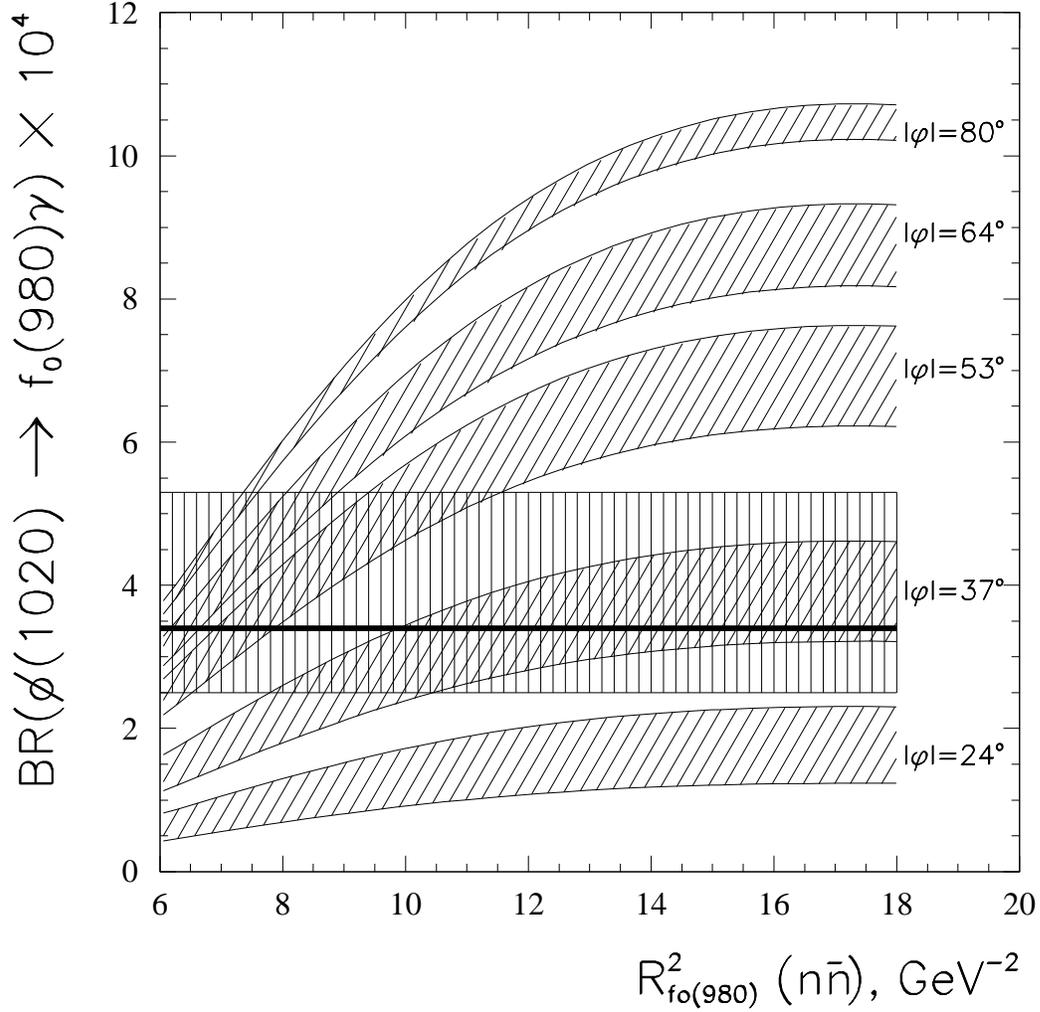,width=15cm}}
\caption{Branching ratio $BR(\phi(1020)\to\gamma f_0(980))$ as a
function of  radius squared of the $n\bar n$ component in
$f_0(980)$. The band with vertical shading stands for the experimental
magnitude:
${\rm BR}(\phi\to\gamma f_0(980))=(3.4\pm 0.4
^{+1.5}_{-0.5})\times 10^{-4}$.
Five other bands, with skew shading, correspond to
$|\varphi|=24^\circ,37^\circ, 53^\circ, 64^\circ, 80^\circ$ at
$-8^\circ\le\varphi_V\le 8^\circ $.}
\end{figure}

\begin{figure}
\centerline{\epsfig{file=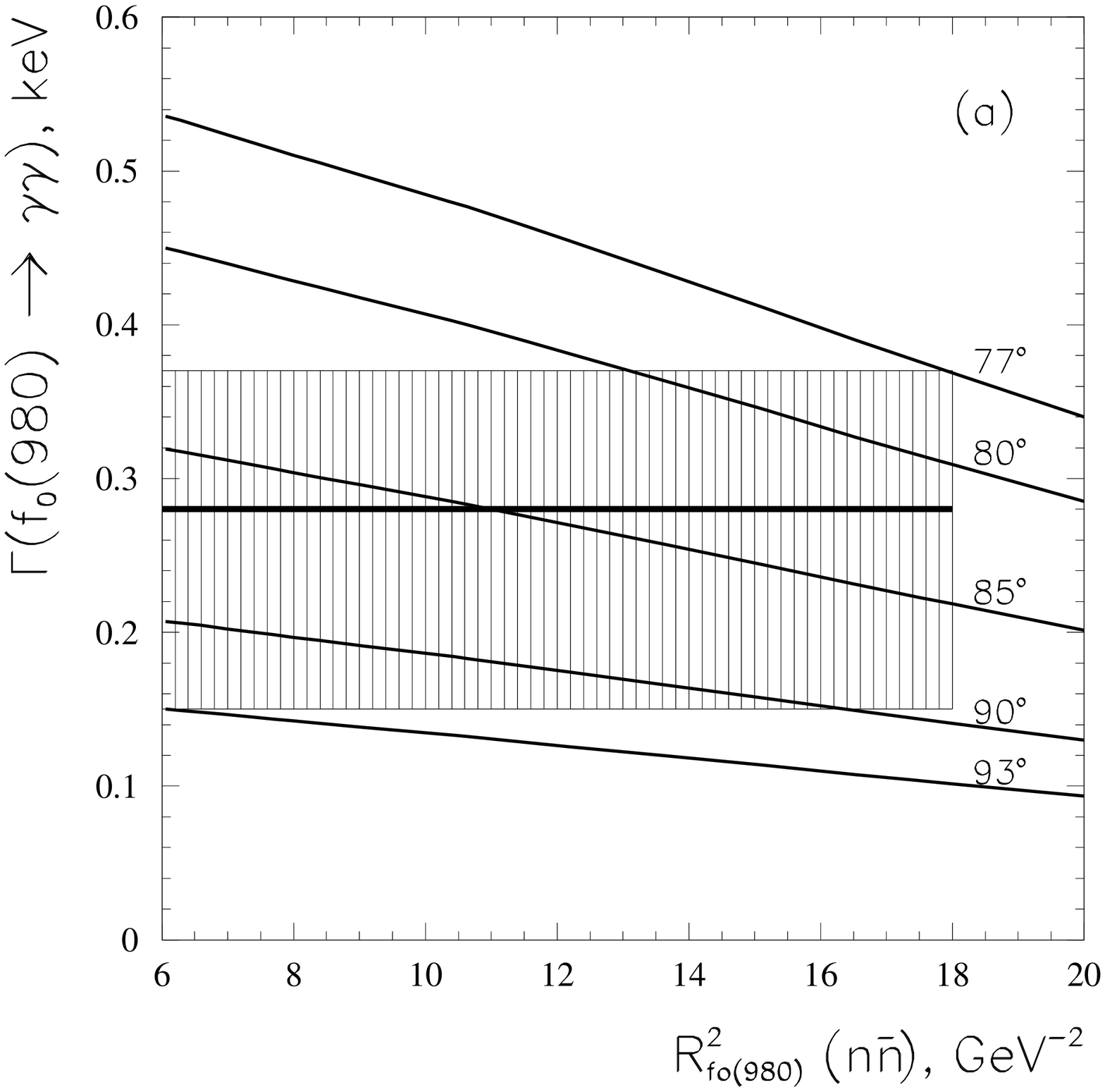,width=8cm}\hspace{1cm}
            \epsfig{file=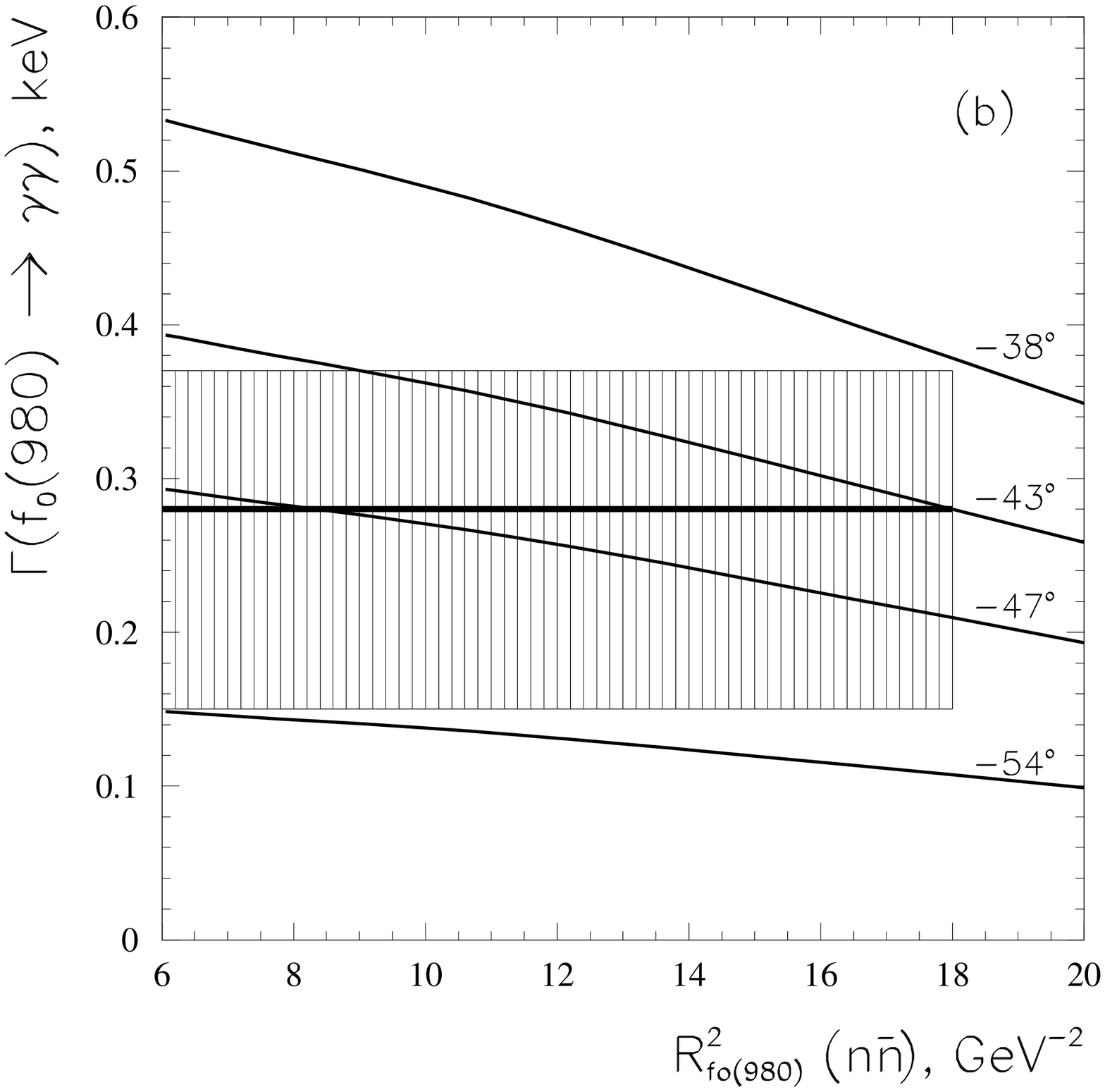,width=8cm}}
\caption{Partial width $\Gamma_{f_0(980) \to \gamma\gamma}$;
experimental data are from [15] (shaded area). a) Curves are
calculated for positive mixing angles
$\varphi=77^\circ,80^\circ,85^\circ,90^\circ,93^\circ$ and b) negative angles
$\varphi=-38^\circ,-43^\circ,-47^\circ,-54^\circ$.}
\end{figure}

\begin{figure}
\centerline{\epsfig{file=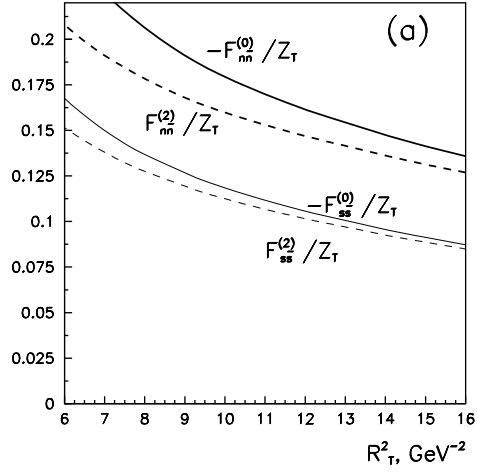,width=7cm}}
\vspace{-0.5cm}
\centerline{\epsfig{file=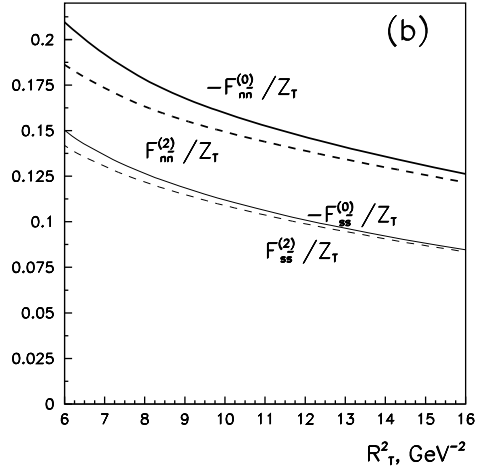,width=7cm}}
\vspace{-0.5cm}
\centerline{\epsfig{file=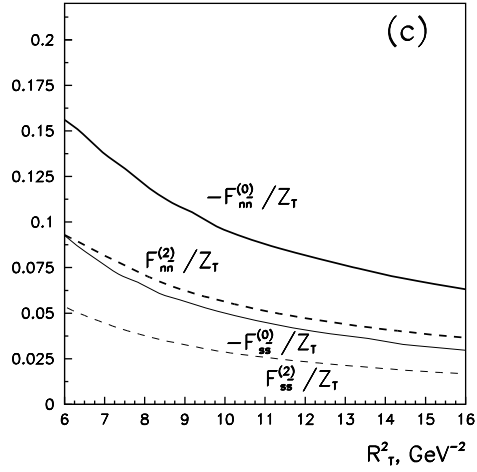,width=7cm}}
\caption{Transition form factors $T\to \gamma\gamma$ (see (17)
or (24)) for the non-strange ($n\bar n$) and strange ($s\bar s$)
quarks versus mean tensor meson radius squared, $R^2_T$.  a)
$F^{(0)}_{q\bar q}(0,0)$ and
$F^{(2)}_{q\bar q}(0,0)$ for $1^3P_2 q\bar q$ state with minimal
vertex, Eq. (41); b) the same as Fig. 6$a$ but with the
vertex determined by (45); c) $F^{(0)}_{q\bar q}(0,0)$
and $F^{(2)}_{q\bar q}(0,0)$ for $1^3F_2 q\bar
q$ state with vertex given by (46).}
\end{figure}

\begin{figure}
\centerline{\epsfig{file=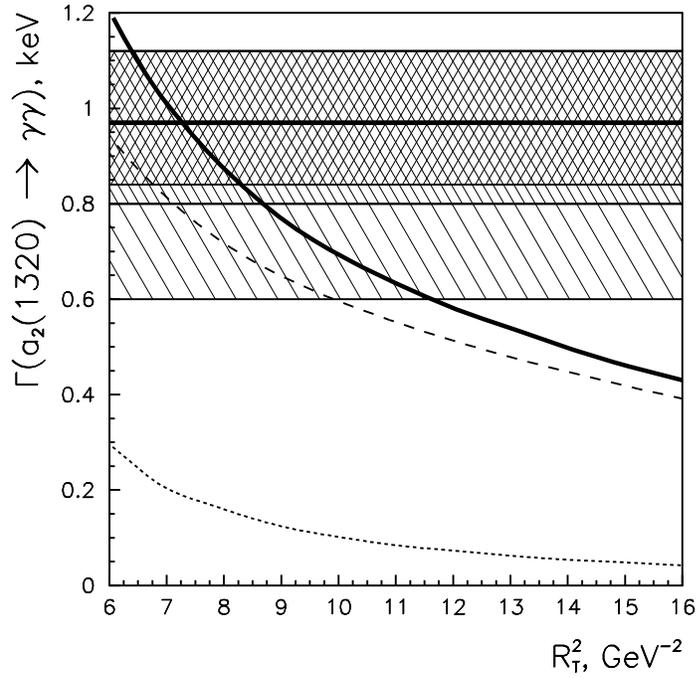,width=10cm}}
\caption{Partial widths for $a_2\to
\gamma\gamma$
versus mean tensor-meson radius squared, $R^2_T$.
Thick solid line: $\Gamma (a_2(1320)\to \gamma\gamma)$
for the vertex given by (43);
Dashed line: $\Gamma (a_2(1320)\to \gamma\gamma)$
for the vertex given by (45)
Dotted line: $\Gamma (a_2(\sim 2000)\to \gamma\gamma)$
for the vertex given by (46).}
\end{figure}

\begin{figure}
\centerline{\epsfig{file=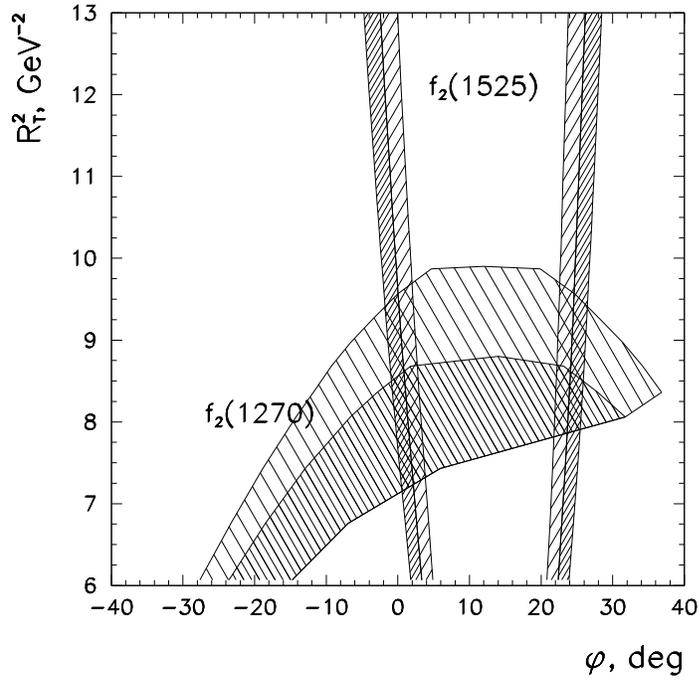,width=10cm}}
\caption{The $(R^2_T,\varphi)$-plot, where $\varphi$ is mixing angle
for  flavour components
$f_2(1270)=n\bar n\cos\varphi  +
 s\bar s\sin \varphi $ and $ f_2(1525)=
- n\bar n\sin\varphi +  s\bar s\cos \varphi $,
with hatched areas which show the regions allowed by data for decays
$f_2(1270)\to \gamma\gamma$ and $f_2(1525)\to \gamma\gamma$.}
\end{figure}

\begin{figure}
\centerline{\epsfig{file=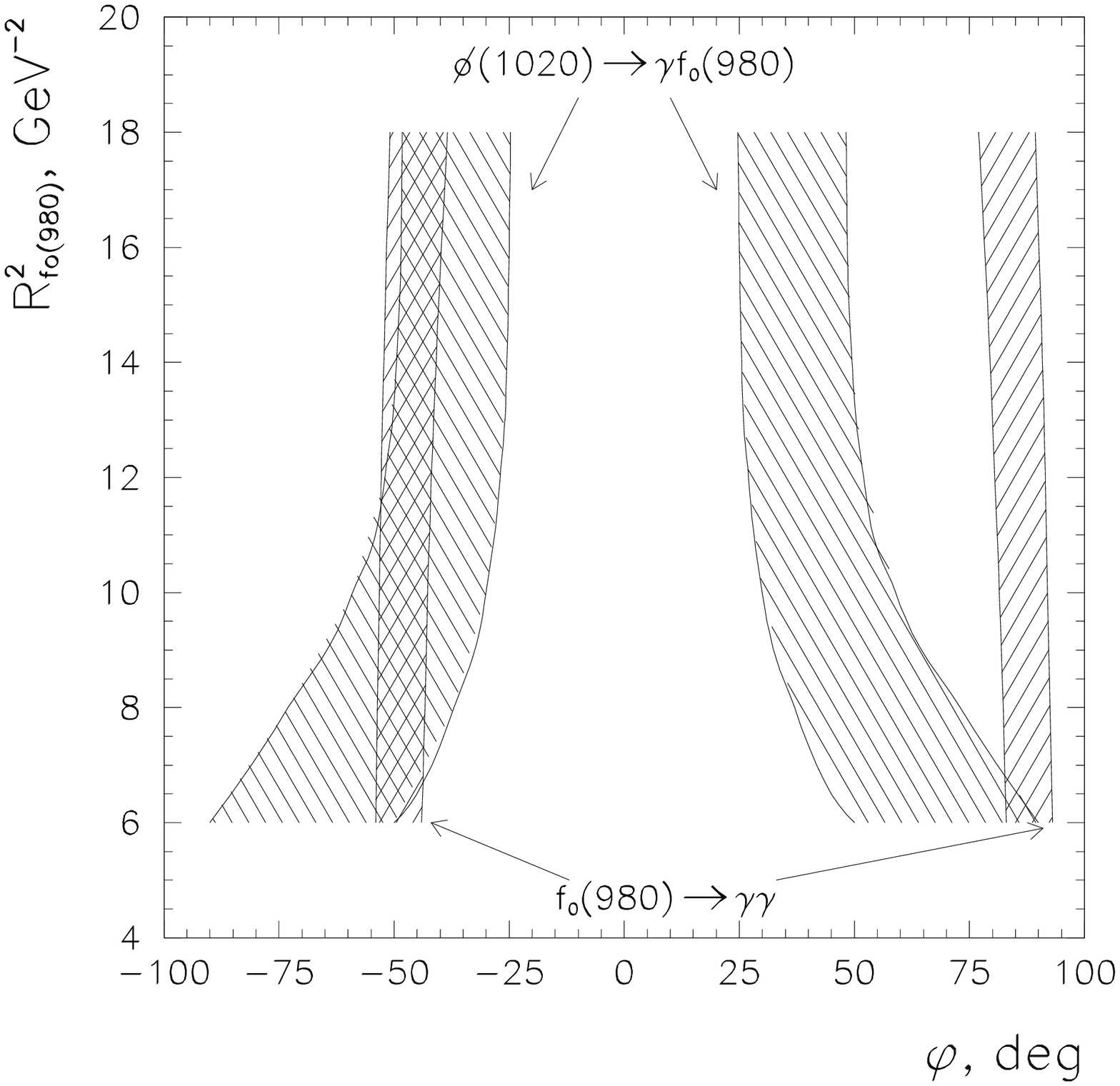,width=15cm}}
\caption{The $(\varphi,R^2_{f_0(980)})$-plot: the shaded areas are
the allowed ones for the reactions
$\phi(1020)\to\gamma f_0(980)$ and $ f_0(980) \to \gamma\gamma$.}
\end{figure}


\begin{thebibliography}{99}
\bibitem{Gatto} R. Gatto, Phys. Lett. {\bf 17}, 124 (1965); \\
V.V. Anisovich, A.A. Anselm, Ya.I. Azimov,
G.S. Danilov and I.T. Dyatlov,
Pis'ma ZETF {\bf 2}, 109 (1965).
\bibitem{klempt}E. Klempt, {\it Meson spectroscopy: glueballs, hybrids,
and $q$ anti-$q$ mesons}, hep-ex/0101031 (2001).
\bibitem{montanet} L. Montanet, Nucl. Phys. Proc. Suppl. {\bf 86}, 381
(2000).
\bibitem{petry} R. Ricken, M. Koll, D. Merten, B.C. Metsch and
H.R. Petry, Eur. Phys. J. A {\bf 9}, 221 (2000).
\bibitem{pnpi}
V.V. Anisovich, A.A. Anselm, Ya.I. Azimov,
G.S. Danilov and I.T. Dyatlov,
Phys. Lett. {\bf 16}, 194 (1965);\\
W.E. Tirring, Phys. Lett. {\bf 16}, 335 (1965);\\
L.D. Solovjev, Phys. Lett. {\bf 16}, 345 (1965);\\
C. Becchi and G. Morpurgo, Phis. Rev. {\bf 140}, 687 (1965).
\bibitem{AKMS}
V.V. Anisovich, M.N. Kobrinsky, D.I. Melikhov and
A.V. Sarantsev, Nucl. Phys. A {\bf 544}, 747 (1992).
\bibitem{pi} V.V. Anisovich, D.I. Melikhov and V.A. Nikonov,
Phys. Rev. D {\bf 52}, 5295 (1995).
\bibitem{eta} V.V. Anisovich, D.I. Melikhov and V.A. Nikonov,
Phys. Rev. D {\bf 55}, 2918 (1997).
\bibitem{f0gg}
A.V. Anisovich, V.V. Anisovich, D.V. Bugg and V.A. Nikonov,
Phys. Lett. B {\bf 456}, 80 (1999).
\bibitem{melikhov}
D.I. Melikhov, Phys. Rev. D {\bf 56}, 7089 (1997);\\
D.I. Melikhov and B. Stech, Phys. Rev. D {\bf 62}:014006 (2000).
\bibitem{novosib} CMD-2 Collaboration: R.R. Akhmetshin {\it et al.},
Phys.  Lett. B {\bf 462}, 371 (1999); {\bf 462}, 380 (1999);\\ SND
Collaboration: M.N. Achasov {\it et al.}, Phys. Lett. B {\bf 485}, 349
(2000).
\bibitem{PDG-00}PDG Group, D.E. Groom {\it et al.},
Eur. Phys. J. C {\bf 15}, 1 (2000).
\bibitem{ABMN} V.V. Anisovich, D.V. Bugg, D.I. Melikhov and V.A. Nikonov,
Phys. Lett. B {\bf 404}, 166 (1997).
\bibitem{AA}
A.V. Anisovich and V.V. Anisovich, Phys. Lett. B {\bf 467}, 289 (1999).
\bibitem{pennington}
M. Boglione and M.R. Pennington,
Eur. Phys. J. C {\bf 9}, 11 (1999).
\bibitem{PDG-98}PDG Group, C. Caso et al., Eur. Phys. J. {\bf C3},
1 (1998).
\bibitem{moment} A.V. Anisovich, V.V. Anisovich, V.N. Markov,
M.A. Matveev and A.V. Sarantsev, {\it Moment-operator expansion
for the two-meson, two-photon and fermion-antifermion states},
hep-ph/0105330 (2001).
\bibitem{spin} V.V. Anisovich, {\it Elements of scattering theory},
Chapter V, in: "Hadron Spectroscopy and Confinement Problem",
Ed. D.V. Bugg, NATO ASI Series, Physics Vol. 353, Plenum Press,
New York and London, (1996).
\bibitem{radius}
V.V. Anisovich, D.V. Bugg and A.V. Sarantsev, Phys. Lett. B
{\bf 437}, 209 (1998); Phys. Atom. Nucl. {\bf 62 }, 289 (1999).
\bibitem{GAMS}
Yu.D. Prokoshkin {\it et al.}, Physics-Doklady {\bf 342},
473 (1995);\\
D. Alde {\it et al.}, Z. Phys. C {\bf 66}, 375 (1995).
\bibitem{a0f0} T. Oest et al. (JADE Coll.) Z. Phys. {\bf C47}
 343 (1990).
\bibitem{a0} D. Antreasyan et al. (Crystal Ball Coll.)
Phys. Rev. {\bf D33} (1986) 1847.
\bibitem{l3} L3 Collaboration, M. Acciarri et al.,
Phys. Lett. {\bf B413}, 147 (1997).
\bibitem{argus} ARGUS Collaboration, H. Albrecht et al., Z. Phys.
{\bf C74}, 469 (1997).
\bibitem{argus90} ARGUS Collaboration, H. Albrecht et al., Z. Phys.
{\bf C48}, 183 (1990).

\end{thebibliography}
\end{document}